 \documentclass[a4paper,aps,prd,10pt,preprintnumbers,showpacs,onecolumn,superscriptaddress,nofootinbib,amsmath,amssymb]{revtex4-1}
\usepackage{amssymb}
\usepackage{amsfonts}
\usepackage{amsmath}
\usepackage{graphicx}
\usepackage{dcolumn}
\usepackage{bm}
\usepackage{makeidx}
 \usepackage[english, english]{babel}
\usepackage{color}

\usepackage{graphicx}
\def\be{\begin{equation}}
 \def\ee{\end{equation}}
 \def\bea{\begin{eqnarray}}
 \def\eea{\end{eqnarray}}

\makeindex

\begin{document}

\title{Motion and collision of particles in rotating linear dilaton black hole}
\author{P. A. Gonz\'{a}lez}
\email{pablo.gonzalez@udp.cl}
 \affiliation{Facultad de
Ingenier\'{i}a y Ciencias, Universidad Diego Portales, Avenida Ej\'{e}rcito
Libertador 441, Casilla 298-V, Santiago, Chile.}
\author{ Marco Olivares }
\email{marco.olivaresr@mail.udp.cl}
\affiliation{ Facultad de Ingenier\'ia y Ciencias, Universidad Diego Portales,
Avenida Ej\'ercito Libertador 441, Casilla 298-V, Santiago, Chile.}
\author{Eleftherios Papantonopoulos}
\email{lpapa@central.ntua.gr}
\affiliation{Department of
Physics, National Technical University of Athens, Zografou Campus
GR 157 73, Athens, Greece.}
\author{Yerko V\'{a}squez}
\email{yvasquez@userena.cl}
\affiliation{Departamento de F\'{\i}sica y Astronom\'{\i}a, Facultad de Ciencias, Universidad de La Serena,\\
Avenida Cisternas 1200, La Serena, Chile.}
\date{\today}


\begin{abstract}
We study the motion of particles in the background of a  four-dimensional linear dilaton black hole. We solve analytically the equations of motion of the test particles and we describe their motion. We show that the dilaton black hole acts as a particle accelerator by analyzing the energy in the center of mass (CM) frame of two colliding particles in the vicinity of its horizon. In particular we find that there is a critical value of the  angular momentum, which depends on the string coupling, and a particle with this critical angular momentum can reach the inner  horizon with an arbitrarily high  CM energy. This is known as the Ba\~nados, Silk and West (BSW) process.  We also show that the motion and collisions of particles have a similar behavior to  the three-dimensional BTZ black hole. In fact, the photons can plunge into the horizon or escape to infinity, and they can not be deflected, while for massive particles there are no confined orbits of first kind, like planetary or circular orbits.
\end{abstract}

\maketitle

\tableofcontents

\section{Introduction}
\label{Introduction}

Stringy black holes are interesting because they have quite different properties from those
that appear in the black hole solutions of general relativity (GR). The main ingredient of these
solutions is the presence of a scalar field, the dilaton, which effectively acts as a string coupling.
Then with a suitable conformal transformation this field couples to the electromagnetic field and this
new coupling alters the known properties of spherically symmetric black hole solutions of GR.
A static stringy charged black hole solution
was found in \cite{Gibbons:1987ps}, and independently in
 \cite{Garfinkle:1990qj}, known as Garfinkle-Horowitz-Strominger (GHS) black hole. These solutions resulted from a four-dimensional field theory of a heterotic string theory.
Then, in the context of  the heterotic string theory, a general electrically charged, rotating black hole solution was found \cite{Sen:1992ua,Sen:1994eb}
which later was extended \cite{Hassan:1991mq} and black hole solutions were found carrying electric charge, and both,
electric and magnetic type antisymmetric tensor gauge field charge. Also, solitonic black hole solutions \cite{Khuri:1997gw} and black hole  solutions in the presence of a cosmological constant \cite{Gao:2005xv} were found.

In \cite{Clement:2002mb}  within the Einstein-Maxwell-dilaton-axion
theory  rotating black hole solutions  which asymptote to the linear dilaton space-time were found. The non-rotating case
can be considered as some limit of extremal
GHS black holes with a geometric interpretation of a horizon plus a throat
 describing a non-asymptotically flat
black hole. Particularly they can
 be regarded as
excitations over the linear dilaton vacua, forming a two-parameter
family, which can be obtained as near-horizon and near-extreme limits
of the parent
dilaton black holes. Also in \cite{Clement:2002mb}  a rotating version  of linear
dilaton black holes were found in which the dilaton field is  independent of the black hole mass.
The metric possesses an ergosphere outside the event horizon, which
rotates with some angular velocity $\Omega_h$. The angular momentum
$J$ computed as a surface integral at spatial infinity contains both
a geometrical contribution and that coming from the Maxwell field.
The Hawking temperature and the entropy depend both on the mass and
the angular momentum of the black hole, these quantities are shown to satisfy
the first law in the form $d{\cal M}=TdS+\Omega_hdJ$. Also analytical solutions of rotating linear dilaton black holes were discussed in \cite{Sakalli:2016fif,Sakalli:2016xoa}.

The Hamilton-Jacobi equation in the rotating linear dilaton black
hole spacetime is shown to be separable \cite{Clement:2002mb}  as in the Kerr case,
indicating the
existence of a Stachel-Killing tensor. Timelike geodesics do not
escape to infinity, while null ones do so for an infinite affine
parameter. The Klein-Gordon equation is also separable, the mode
behavior near the horizon exhibits the superradiance phenomenon.
It was shown that all
superradiant modes are reflected at large distances from the black
hole, so there is no superradiant flux at infinity, in contrast with
the case of massless fields in the Kerr spacetime. This situation
is similar to the confinement of {\em massive}
superradiant modes in the Kerr metric, which are reflected back to the
horizon causing stimulated emission and absorption. In the
classical limit this leads to an amplification effect due to
the positive balance between emission and absorption.
In the Kerr spacetime this phenomenon is rather small, being present
only for massive modes. In our case massless modes are confined too,
therefore all superradiant modes will form a cloud outside
the horizon with an exponentially growing amplitude.
This is in fact a classical instability
which manifests itself in the rapid transfer of angular momentum
from the hole to the outside matter cloud. A similar conclusion
was obtained in \cite{Hawking:1999dp} for the case of the Kerr-AdS spacetime
with reflecting boundary conditions on the AdS boundary. The stability  of the rotating linear dilationic black hole was also studied in \cite{Li:2012zj}, by calculating the quasinormal modes of scalar field perturbations and  superradiant
modes.

The properties of charged black holes in string theory can be revealed studying the geodesics around these solutions. This is because except the information we get solving the classical equations of motion in the form of Einstein equations we also get information about stringy corrections due to the string coupling which is of the order of Planck scale. The study of null geodesics in the electrically charged GHS black hole was carried out in Refs.  \cite{Fernando:2011ki, Soroushfar:2016yea}, and the timelike geodesics were analyzed in   \cite{Soroushfar:2016yea, Maki:1992up, Olivares:2013jza, Blaga:2014lva, Blaga:2014spa}. Additionally, in \cite{Villanueva:2015kua} the gravitational Rutherford scattering and Keplerian orbits were studied in the GHS black hole background and  the motion of massive particles with electric and magnetic charges in the background of a magnetically charged GHS stringy black hole in Ref. \cite{Gonzalez:2017kxt}.

The aim of this work is to consider the linear dilaton black hole solution \cite{Clement:2002mb}, and to study the geodesic structure. Also, to analyze, via the Ba\~nados, Silk and West (BSW) mechanism, the possibility of obtaining unbounded energy in the center of mass (CM) frame of two colliding particles. Two particles colliding near the degenerate horizon of an extreme Kerr black hole could create a large CM energy if one of the particles has a critical angular momentum; thus, extreme Kerr black holes can act as natural particle accelerators, which is known as BSW mechanism  \cite{Banados:2009pr}.  Then, based on
an infinite acceleration being able to occur not only for extremal black holes but also for non-extremal ones \cite{Grib:2010at}, it was shown that an infinite energy in the CM frame of colliding particles is a universal property of rotating black holes as long as the angular momentum of one of the colliding particles approaches the critical value \cite{Zaslavskii:2010jd}.
However, it was shown that the similar effect exists for non-rotating charged black holes
\cite{Zaslavskii:2010aw}. Moreover, the extension of the BSW mechanism to non-extremal backgrounds shows that particles cannot collide with arbitrarily high energies at the outer horizon and that ultra-energetic collisions  can only occur near the Cauchy horizon of a Kerr black hole with any spin parameter \cite{Gao:2011sv}. The non-extremal Kerr-de Sitter black holes could also act as particle accelerators with arbitrarily high CM energy if one of the colliding particles has the critical angular momentum \cite{Li:2010ej}. Also,
it is known that two particles in the ergosphere lead to infinity growth of the energy of the CM frame, provided the angular momentum of one of the two particles has a large negative angular momentum and a fixed energy at infinity for the Kerr black holes \cite{Grib:2010at},
which was subsequently proven to be a
universal property of the ergosphere \cite{Zaslavsky:2013dra}. The BSW mechanism has been studied for different black hole geometries.
Higher-dimensional
 black holes
have been studied, for instance,
in Refs. \cite{Abdujabbarov:2013qka, Debnath:2015bna, Zaslavskii:2016stw}, lower-dimensional black holes
in Refs. \cite{Sadeghi:2013gmf, Lake:2010bq, Yang:2012we, Tsukamoto:2017rrl,Fernando:2017kut, Becar:2017aag}, as well as  stringy black holes  \cite{Fernando:2013qba}.

The manuscript is organized as follows: In Sec. \ref{background} we give a brief review of the linear dilaton  black hole. Then, we obtain the equations of motion in Sec. \ref{STL}. In Sec. \ref{GEP} we analyze the geodesics in the equatorial plane.
In Sec. \ref{CMS} we obtain the CM energy of two colliding particles, and we study the radial motion of a particle with critical angular momentum and we investigate the possibility that
the black hole acts as a particle accelerator. Then, in Sec \ref{G4D} we study the geodesics in four dimensions.
Finally, our conclusions are in Sec. \ref{conclusion}.

\section{Static linear dilaton black holes}
\label{background}

In this section we will review how the linear dilaton black hole is extracted from the  GHS solution \cite{Garfinkle:1990qj}. The most general action of low energy heterotic string theory is given by
\be
\mathcal{S}= \int d^D x \sqrt{-g}\ e^{-2\varphi}
 \Big{[}
\Lambda +R+4(\nabla \varphi)^2 -
F_{\mu\nu}F^{\mu\nu} - \frac{1}{12} H_{\mu\nu\rho}
H^{\mu\nu\rho}
 \Big{]}
~, \label{action}
\ee
where the scalar field $\varphi$ is the dilaton field, $F_{\mu\nu}$ is a Maxwell field,
and the three form $H_{\mu\nu\rho}$ is  related to
a two-form potential  $B_{\mu\nu}$  and the gauge field $A_\mu$
by $H = dB - A \wedge F$
so that $dH = - F\wedge F $. Note that in this action the term $e^\varphi$
plays the role of a coupling constant giving the strength of the stringy effects.

If we set $H$ to zero and make  the conformal transformation of the metric
to rescale $g_{\mu\nu}$ by
$e^{-2\varphi}$ to get a metric with the standard Einstein action \be g^E_{\mu\nu} =
e^{-2\varphi} g_{\mu\nu}~. \ee  The action now becomes (with $\Lambda=0$)
\be
\mathcal{S}=\int d^4x\ \sqrt{-g_E}\ \left(R_E - 2(\nabla\varphi)^2 - e^{-2\varphi}
F^2\right)~. \label{simaction}
\ee

Rescaling  the Einstein metric we get a simple  form of an electrically charged black hole
\be
ds^2_E = - (1-\frac{2M}{r}) dt^2 + (1-\frac{2M}{r})^{-1} dr^2
+ r(r-\frac{Q^2}
{M} )d\Omega~, \label{swsol} \ee
\be F_{rt}=\frac{Q}{r^2}~, \qquad e^{2\phi}
= 1-\frac{Q^2}{Mr}~. \label{swfil} \ee
Note that the metric in the $r-t$
plane is identical to Schwarzschild. The only difference is
that the area
of the spheres is smaller.
In fact, this area goes to zero when $r=Q^2/M$ and
this surface is singular. Since $g_{tt}$ remains finite at the singularity,
there
is no ``infinite stretching" analogous to what happens to an observer hitting
the singularity in Schwarzschild.
As you increase $Q$, the singularity moves out in ``$r$''.  In the extremal
limit $Q^2 = 2M^2$, the singularity coincides with the horizon.
If you increase $Q$ farther, the
singularity moves outside the horizon and becomes timelike.

Using these properties of the electrically charged GHS black hole, studying its  near-horizon and near-extremal limits and the results of \cite{Giddings:1992kn} a
linear dilaton black hole solution was constructed in \cite{Clement:2002mb}
\begin{eqnarray}
ds^2&=&\frac{r-b}{r_0}dt^2-\frac{r_0}{r-b} dr^2-r_0 r d\Omega^2,
\label{gstatic}\\
e^{2\phi}&=&\frac{r}{r_0}, \qquad
F_{tr}=\frac{1}{\sqrt{2}r_0}~. \label{phiFstatic}
\end{eqnarray}
The two parameters $r_0$ and $b$ have a different physical
meanings. The first describes the electric charge $Q$ of
the solution which is given
by the flux through a 2-sphere
\begin{equation}\label{Q}
Q=\frac1{4\pi}\int e^{-2\phi} F^{0r}\sqrt{g}\,d\Omega=\frac{r_0}{\sqrt{2}}~,
\end{equation}
and characterizes the linear dilaton background and the
parameter $b$ characterizing the  black hole solution is proportional to its mass.
This solution corresponds to the
Schwarzschild black hole. However the spacetime is not
asymptotically flat.

Finally, this static lineal dilaton  solution was generalized to a stationary one \cite{Clement:2002mb}
\begin{eqnarray}
ds^2 & = & \frac{r^2-2Mr+a^2}{r_0 r}dt^2-r_0
r\left[\frac{dr^2}{r^2-2Mr+a^2}
+d\theta^2+\sin^2\theta\bigg(d\varphi-\frac{a}{r_0 r}dt
\bigg)^2\right]\,, \label{nkgN0} \\
F & = &
\frac1{\sqrt{2}}\bigg[\frac{r^2-a^2\cos^2\theta}{r_0r^2}\,dr\wedge dt
+ a\sin 2\theta\,d\theta\wedge\bigg(d\varphi-\frac{a}{r_0 r}dt
\bigg)\bigg]\,, \label{nkFN0} \\
e^{-2\phi} & = & \frac{r_0r}{r^2+a^2\cos^2\theta}  \,.\label{nkdaN0}
\end{eqnarray}
The Maxwell two-form is derivable from the following four-potential
\begin{equation}\label{pots}
A=\frac1{\sqrt{2}}\left(\frac{r^2+a^2\cos^2\theta}{r_0 r}dt+
a\sin^2\theta\,d\varphi\right),
\end{equation}
where the gauge is chosen such that the vector magnetic potential be
regular on the axis $\theta = 0$.

Although the metric (\ref{nkgN0}) was derived from the Kerr metric, it differs
in its asymptotic behavior ($r \to \infty$), which is the same as
for the linear dilaton metric and in its behaviour
near $r = 0$. In the case of the Kerr metric, $r = 0$ is the
equation of a disk through which the metric can be continued to
negative $r$, while in (\ref{nkgN0}) $r = 0$ is a timelike line
singularity. It follows studying  the Penrose diagrams of
(\ref{nkgN0}) for the three cases  $M^2>a^2$, $M^2=a^2$ and
$M^2<a^2$ are identical, not to those of the Kerr spacetime, but
rather to those of the Reissner-Nordstr\"{o}m spacetime, with the
charge replaced by the angular momentum parameter $a$.

\section{Motion equations of particles in a linear dilaton black hole}
\label{STL}
Now, in order to obtain the motion equations of particles and photons in a linear dilaton black hole we rewrite the metric (\ref{nkgN0}) as
\begin{equation}
ds^2=-f(r)dt^2+\frac{dr^2}{f(r)}+h(r) \left( d\theta^2+\sin ^2 \theta \left(  d \varphi -a \frac{dt}{h(r)}\right)^2 \right)~,
\end{equation}
where
\begin{equation}
f(r)=\frac{r^2-2Mr+a^2}{r_{0} r}=\frac{(r-r_+)(r-r_-)}{r_0 r}, \,\,\, h(r)=r_0 r~,
\end{equation}
and $r_0$ is a positive constant. The outer and inner horizons are given by
\begin{equation}\label{rp}
r_{\pm}=M\pm \sqrt{M^2-a^2}~.
\end{equation}
It is worth noting that the geodesic structure of the linear dilaton black hole was studied in \cite{Hamo:2015ica} for the non-rotating case. Here, we shall extend such study by considering the rotating case.

So,
we first derive the equations of motion following the same approach given in \cite{chandra}. We consider the motion of test particles with mass $m$ and photons. Thus, in order to obtain the equation of motion, we apply the Hamilton-Jacobi formalism. In this sense, the Hamilton-Jacobi equation  for the geometry described by the metric $g_{\mu\nu}$ is
\begin{equation}
2\frac{\partial S}{\partial\tau}=
g^{\mu \nu}\frac{\partial S}{\partial
x^{\mu}}\frac{\partial S}{\partial
x^{\nu}}~.\label{i.3}
\end{equation}
So, in order to solve the Hamilton-Jacobi equation, and taking into account the symmetries of the metric, we use the following ansatz for the action
\begin{equation}
S=-{1\over2}m^2\tau-E\, t+S_{r}(r)+S_{\theta}(\theta)+L\,\varphi~, \label{i.5}
\end{equation}
where $E$ and $L$ are identified as the energy and angular momentum
of the particle. Thus, by replacing this ansatz in Eq. (\ref{i.3}) yields
\begin{equation}
\left( \frac{\partial S_{r}}{\partial r}\right)^2-\left( \frac{E}{f(r)}-\frac{aL}{h(r) f(r)} \right)^2+\frac{m^2}{f(r)}+\frac{1}{h(r) f(r)} \left( \left( \frac{\partial S_{\theta}}{\partial \theta} \right)^2 +\frac{L^2}{\sin ^2 \theta} \right)=0~.\label{i.6}
\end{equation}
Now, using the standard procedure we recognize the following constant
\begin{equation}
k^2= \left( \frac{\partial S_{\theta}}{\partial \theta} \right)^2 +\frac{L^2}{\sin ^2 \theta}~,\label{i.7}
\end{equation}
which is identified as the Carter separability constant. Then, we obtain the following equation for the radial component of the action
\begin{equation}
\left( \frac{\partial S_{r}}{\partial r}\right)^2-\left( \frac{E}{f(r)}-\frac{aL}{h(r) f(r)} \right)^2+\frac{k^2+m^2 h(r)}{h(r) f(r)}=0~.\label{i.8}
\end{equation}
Thus, the formal solutions for the radial and polar components of the action are given by
\begin{equation}
S_r (r,k^2,E,L)=\epsilon \int \sqrt{\left( E-\frac{aL}{r_0 r}\right)^2-f(r) \left( m^2+ \frac{k^2}{r_0 r}\right)} \frac{dr}{f(r)}~,\label{i.9}
\end{equation}
\begin{equation}
S_{\theta}(\theta, k^2, L) = \epsilon \int \sqrt{k^2-\frac{L^2}{\sin ^2 \theta}} d\theta~,\label{i.10}
\end{equation}
where $\epsilon=\pm 1$.
Now, using the Hamilton-Jacobi method and making
${\delta S\over\delta k^2}=0$,
${\delta S\over\delta m^2}=0$, ${\delta S\over\delta E}=0$ and ${\delta S\over\delta L}=0$, we simplify our study to the following quadrature problem
\begin{eqnarray}
 && \int \frac{d \theta}{\sqrt{k^2-\frac{L^2}{\sin ^2 \theta}}}=\int \frac{dr}{r_{0} r \sqrt{\left( E-\frac{aL}{r_0 r}\right)^2-f(r) \left( m^2+ \frac{k^2}{r_0 r}\right)}}~, \\
  && \tau (r)= \epsilon \int \frac{dr}{\sqrt{\left( E-\frac{aL}{r_0 r}\right)^2-f(r) \left( m^2+ \frac{k^2}{r_0 r}\right)}}~, \\
&&t (r)= \epsilon \int \frac{\left(E-\frac{aL}{r_{0} r}\right)dr}{f(r) \sqrt{\left( E-\frac{aL}{r_0 r}\right)^2-f(r) \left( m^2+ \frac{k^2}{r_0 r}\right)}}~, \\
\label{varphi}
&&\varphi (r)= \epsilon \int \frac{\left(E-\frac{aL}{r_{0} r}\right) \frac{a}{r_{0} r}}{\sqrt{\left( E-\frac{aL}{r_0 r}\right)^2-f(r) \left( m^2+ \frac{k^2}{r_0 r}\right)}}\frac{dr}{f(r)}+\epsilon \int \frac{L}{\sin ^2 \theta \sqrt{k^2-\frac{L^2}{\sin ^2 \theta}}}d\theta~.
\end{eqnarray}
Then, by defining the Mino time $\gamma$ as $d\gamma= d\tau /(r_0 r)$,
we can write the equations of motion in terms of the new time parameter, which yields
\begin{equation}\label{angular}
\left( \frac{d \theta}{d \gamma} \right)^2=k^2-\frac{L^2}{\sin ^2 \theta}~,
\end{equation}
\begin{equation}\label{eqradial}
\left( \frac{d r}{d \gamma} \right)^2=r_0^2 r^2 \left(  \left( E-\frac{aL}{r_0 r}\right)^2-f(r) \left( m^2+\frac{k^2}{r_0 r}\right) \right)~,
\end{equation}
\begin{equation}
\frac{d \varphi}{d \gamma}= \left( E-\frac{aL}{r_0 r}\right) \frac{a}{f(r)}+\frac{L}{\sin ^2 \theta}~,
\end{equation}
\begin{equation}
\frac{d t}{d \gamma} = \frac{r_0 r \left( E-\frac{aL}{r_0 r} \right)}{f(r)}~.
\end{equation}
In this way we have the equations of motion for our particles moving in the background of linear dilaton black hole. In the next section we will perform an analysis of the equations of motion in the equatorial plane.


\section{Geodesics in the equatorial plane}
\label{GEP}
The Lagrangian associated with the motion in the equatorial plane, that is, $\theta=\pi/2$ and $\dot{\theta}=0$ results to be:
\begin{equation}\label{tl4}
  2\mathcal{L}=-{r^2-2Mr\over r_0r}\dot{t}^2-2a\dot{t}\dot{\phi}+
 { r_0r\over r^2-2Mr+a^2}\dot{r}^2+ r_0r\,\dot{\phi}^2~,
\end{equation}
where $\dot{q}=dq/d\tau$, and $\tau$ is an affine parameter along the geodesic that
we choose as the proper time. Since the Lagrangian (\ref{tl4}) is
independent of the cyclic coordinates ($t,\phi$), then their
conjugate momenta ($\Pi_t, \Pi_{\phi}$) are conserved. Then, the equations of motion are obtained from
$ \dot{\Pi}_{q} - \frac{\partial \mathcal{L}}{\partial q} = 0$, and yield
\begin{equation}
\dot{\Pi}_{t} =0 , \quad \dot{\Pi}_{r} =
-{\dot{t}^{2}\over 2\, r_0}-
{ r_0(r^2-a^2)\,\dot{r}^{2}\over 2(r^2-2Mr+a^2)^2}
+{r_0\dot{\phi}^2\over 2}\quad \textrm{and}\quad \dot{\Pi}_{\phi}=0~,
\label{w.11a}
\end{equation}
where $\Pi_{q} = \partial \mathcal{L}/\partial \dot{q}$
are the conjugate momenta to the coordinate $q$, and are given by
\begin{equation}
\Pi_{t} = -{r^2-2Mr\over r_0r} \dot{t} -a\,\dot{\phi}\equiv -E~, \quad \Pi_{r}={ r_0r\over r^2-2Mr+a^2}\dot{r}~\textrm{and}\quad \Pi_{\phi}
=-a \dot{t} +r_0r\,\dot{\phi}\equiv L~,
\label{w.11c}
\end{equation}
where $E$ and $L$ are dimensionless integration constants associated to each of them.
Therefore, the Hamiltonian is given by
\begin{equation}
\mathcal{H}=\Pi_{t} \dot{t} + \Pi_{\phi}\dot{\phi}+\Pi_{r}\dot{r}
-\mathcal{L}
\end{equation}
\begin{equation}
2\mathcal{H}=-E\, \dot{t} + L\,\dot{\phi}+{ r_0r\over r^2-2Mr+a^2}\dot{r}\equiv -m^2~.
\label{w.11z}
\end{equation}
Now, by normalization, we shall consider that $m^2 = 1$ for massive particles and $m^2 = 0$ for photons. Therefore, we obtain
\begin{eqnarray}
\label{w.14}
&&\dot{\phi}= {r_0r\,aE+(r^2-2Mr)L\over r_0r(r^2-2Mr+a^2)}~,\\
\label{w.12}
&&\dot{t}= { r_0rE-aL\over r^2-2Mr+a^2}~,\\
\label{w.13}
&&\dot{r}^{2}=   \left( E-\frac{aL}{r_0 r}\right)^2-{ r^2-2Mr+a^2\over r_0r} \left( m^2+\frac{L^2}{r_0 r}\right)=\left(E-V_-\right)\left(E-V_+\right)~,
\end{eqnarray}
where $V_{\pm}(r)$ is the effective potential and it is given by
\begin{equation}\label{tl8}
  V_{\pm}(r)=\frac{aL}{r_0 r}\pm\sqrt{{ r^2-2Mr+a^2\over r_0r} \left( m^2+\frac{L^2}{r_0 r}\right) }~.
\end{equation}
Since the negative branch have no classical interpretation, it is associated with antiparticles in the framework of quantum field theory \cite{Deruelle:1974zy}, we choose the positive branch of the effective potential $V = V_+$.
In the next section we will perform a general analysis of the equations of motion.

\subparagraph{Dragging of inertial frames:}
In linear dilaton black hole, the presence of $g_{t\phi}\neq 0$
in the metric introduces qualitatively new effects on particle and photon trajectories,
like the effect  called ``dragging of inertial frames''. Thus,
 by using the equations (36) and (37), we obtain
\begin{equation}
\dot{\phi}/ \dot{t}={d\phi \over dt}={r_0aE+(r-2M)L \over r_0(r_0rE-aL)}
\equiv \omega (r)~.
\end{equation}
This equation defines $\omega (r)$, the angular velocity of the test particle. Notice that  if we consider  a zero angular-momentum particle $(L=0)$, the angular velocity of the test particle is given by
\begin{equation}
\omega (r)={a \over r_0\,r}~.
\end{equation}
So, a remarkable result is that a particle dropped ``straight in'' ($L=0$) from a finite distance is ``dragged'' just by the influence of gravity. Therefore, the test particle acquires an angular velocity  ($\omega$), with the same sign that $a$. Also, this effect weakens with distance as $1/r$. It is worth to mention that for the Kerr metric this effect weakens with the distance as $1/r^3$ \cite{shutz}.

\subparagraph{Ergoregion:}
Following a similar treatment given in Ref. \cite{shutz} for the Kerr geometry, we consider photons emitted in the equatorial plane $(d\theta=0)$ at some given $r$. In particular, we consider those initially going in the $\pm\phi$-direction, that is tangent to a circle of constant $r$, $(dr=0)$. Then, they generally have only $dt$ and $d\phi$ nonzero on the path at first and since $ds^2 = 0$, from the metric (2.9), we have
\begin{equation}\label{eq11}
0=g_{tt} \,dt^2-2\,a \sin^2\theta \,dt \, d\phi+r_0\,r\sin^2\theta \,d\phi^2~,
\end{equation}
where
\begin{equation}
g_{tt}={r^2-2Mr+a^2\cos^2\theta \over r_0\,r}~.
\end{equation}
Thus, Eq. 	(\ref{eq11}) can be written as
\begin{equation}
0=g_{tt} -2\,a \sin^2\theta  \, \omega+r_0\,r\sin^2\theta \,\omega^2~.
\end{equation}
It is worth to mention that a remarkable effects happens if $g_{tt}=0$, the two solutions are given by
\begin{equation}
\omega_1=0 \quad \text{and} \quad \omega_2={2a\over r_0\,r}~.
\end{equation}
Notice that the second solution, $\omega_2$,  gives  the same sign as the parameter $a$, and it represents the photon sent off in the same direction as the rotating hole. The other solution, $\omega_1$,  means that the other photon the one sent backwards initially doesn't move at all. The dragging of orbits has become so strong that this photon cannot move in the direction opposite to the rotation. The surface where $g_{tt}=0$ is called the ergosphere or ``static limit'', since inside it no particle can remain at fixed $ r, \theta, \phi$, and this occurs at
\begin{equation}
r_{e}=M+\sqrt{M^2-a^2\cos^2\theta}~,
\end{equation}
being $r_e$ the radius of the ergosphere, which lies outside the horizon (\ref{rp}) and coincides with the event horizon only at the poles $\theta=0$ and $\theta=\pi$. Inside this radius, since $g_{tt}>0$, all particles and photons must rotate with the black hole. This result is similar to the Kerr metric  \cite{shutz}.

On the other hand, as in the Kerr solution, in the linear dilaton black hole the ergosphere occurs at $g_{tt}=0$ and the horizon is at $g_{rr}=0$. The horizon $r_+$, is a surface of constant $r$ and $t$, any surface of constant $r$ and $t$ has an intrinsic metric whose line element comes from Eq. (2.9) with $dt = dr = 0$:
\begin{equation}
ds^2=r_0\,r_+ \,d\theta^2+r_0\,r_+\sin^2\theta \,d\phi^2~.
\end{equation}
The proper area of this surface, $\mathcal{A}$,  is given by integrating the square root of the determinant of this metric over all $\theta$ and $\phi$:
\begin{equation}
\mathcal{A}=\int_{0}^{2\pi}\int_{0}^{\pi} r_0\,r\sin\theta\,d\theta\,d\phi=4\pi \,r_0r_+~,
\end{equation}
which is different to the result for the Kerr geometry \cite{shutz}.

\subsection{Time Like Geodesics }

\begin{figure}[!h]
	\begin{center}
		\includegraphics[width=80mm]{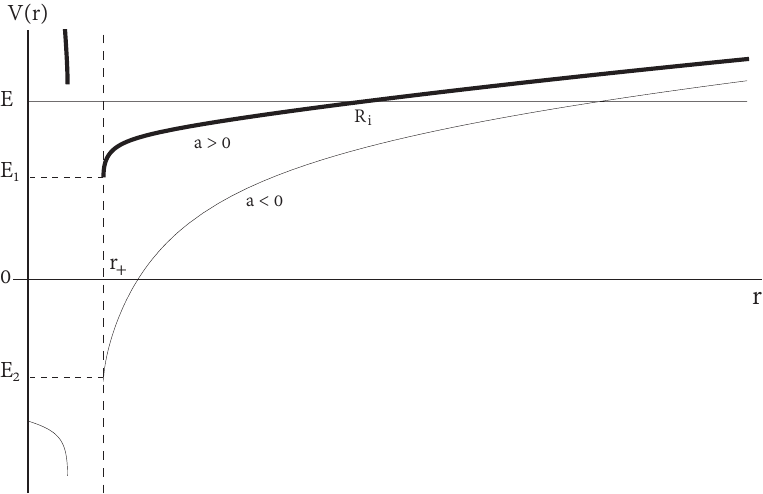}
	\end{center}
	\caption{The behavior of the effective potential for massive particles as a function of $r$, $r_0=1$, $M=2$, $L=4$ and $\left| a\right| =1.9$. }
	\label{fig.1}
\end{figure}

In this case $(m=1)$, is adequate to rewrite the effective potential in the
following form
\begin{equation}\label{tl81}
V(r)=\frac{aL}{r_0 r}+\sqrt{{ r^2-2Mr+a^2\over r_0r} \left( 1+\frac{L^2}{r_0 r}\right) }~,
\end{equation}
whose behaviors is given in Fig. \ref{fig.1} for particles with
$a>0$ (direct geodesics) and $a<0$ (retrograde geodesics), where $E$ corresponds to the energy of the particle, whose trajectory is allow for direct and retrograde geodesics,  $E_1= V(r_+, \left| a\right|)$ and $E_2= V(r_+, -\left| a\right|)$. Notice that retrograde geodesics allow trajectories with negative energy with $E>E_2$. However, the behavior of the trajectories are similar in both cases, being bounded orbits with a finite return point $R_i$, which is given by
\begin{equation}
R_i= {r_0^2E^2+2Mr_0-L^2\over 2\,r_0}\left( 1+\sqrt{1+{4r_0(2ML^2-2aELr_0-r_0a^2) \over (r_0^2E^2+2Mr_0-L^2)^2} }\right)~.
\label{ta2}
\end{equation}
This return point for a given energy $E$ is greater for retrograde geodesics than for direct geodesics. Now, in oder to obtain the solution for the proper time we integrate Eq. (\ref{w.14}), and we obtain
\begin{equation}
\tau(r)=\frac{r_0(F(r)-F(R_i))}{4\sqrt{2ML^2-2aELr_0-r_0a^2}}~,
\label{tr4}
\end{equation}
where
\begin{equation}
F(r)=\frac{2\zeta(\Omega)\wp^{-1}[U(r)]}{\wp^{'}(\Omega)}
+\frac{1}{\wp^{'}(\Omega)}\ln\left| \frac{\sigma[\wp^{-1}[U(r)]-\Omega]}
{\sigma[\wp^{-1}[U(r)]+\Omega]}\right|~,
\label{tr4}
\end{equation}
being $\wp^{-1}(Y)\equiv\wp^{-1}(Y; g_2, g_3)$ the inverse
P-Weierstrass function, $\sigma(Y)\equiv \sigma(Y; g_2, g_3)$ is the sigma Weierstrass function,
$\zeta(Y)\equiv \zeta(Y; g_2, g_3)$ is the zeta Weierstrass function,
and the Weierstrass invariants are given by
\begin{equation}
g_2={1\over 12}\left({ r_0^2E^2+2Mr_0-L^2\over 2ML^2-2aELr_0-r_0a^2} \right)^2
+{1\over 4}\left({ r_0\over 2ML^2-2aELr_0-r_0a^2} \right)~,
\label{tr5}
\end{equation}
and
\begin{equation}
g_3={-1\over 216}\left({ r_0^2E^2+2Mr_0-L^2\over 2ML^2-2aELr_0-r_0a^2} \right)^3
-{1\over 48}{r_0( r_0^2E^2+2Mr_0-L^2)\over (2ML^2-2aELr_0-r_0a^2)^2}~.
\label{tr5}
\end{equation}
Also, the radial function $U=U(r)$ and the constant $\Omega$ are given by
\begin{equation}
\label{tr6.1}
U(r)={1 \over 4r}+{1\over 12}\left({ r_0^2E^2+2Mr_0-L^2\over 2ML^2-2aELr_0-r_0a^2} \right)~,
\end{equation}
and
\begin{equation}
\Omega=\wp^{-1}\left({1\over 12}\left({ r_0^2E^2+2Mr_0-L^2\over 2ML^2-2aELr_0-r_0a^2} \right)\right)~.
\label{tr6.2}
\end{equation}
Now, we consider the solution of the trajectory in the coordinate time in order to compare both times
\begin{equation}
t(r)=-A_+\left[F_+(r)-F_+(R_i)\right]+A_-\left[F_-(r)-F_-(R_i)\right]~,
\end{equation}
where
\begin{equation}
F_{\pm}(r)=\frac{2\zeta(\Omega_{\pm})\wp^{-1}[U(r)]}{\wp^{'}(\Omega_{\pm})}
+\frac{1}{\wp^{'}(\Omega_{\pm})}\ln\left| \frac{\sigma[\wp^{-1}[U(r)]-\Omega_{\pm}]}
{\sigma[\wp^{-1}[U(r)]+\Omega_{\pm}]}\right|
~,\label{tr4}\end{equation}
\begin{equation}
 \Omega_{\pm}=\wp^{-1}\left[{1\over 12}\left({ r_0^2E^2+2Mr_0-L^2\over 2ML^2-2aELr_0-r_0a^2} \right)+{1\over 4r_{\pm}}\right]~,
\label{tr6}
\end{equation}
\begin{equation}
A_{\pm}={r_0(r_0E-aLr_{\pm})\over 4(r_+-r_-)\sqrt{2ML^2-2aELr_0-r_0a^2}}~.
\label{tr6}
\end{equation}
Then, in Fig. \ref{f4.1} we plot  the proper $(\tau)$ and coordinate time $(t)$ as function of $r$ for a particle falling from a finite distance with zero initial velocity, we can see that the particle falls towards the horizon in a finite proper time. The situation is very different if we consider the trajectory in the coordinate time, where t goes to infinity. This physical result is consistent with the Kerr black hole.
\begin{figure}[!h]
	\begin{center}
		\includegraphics[width=80mm]{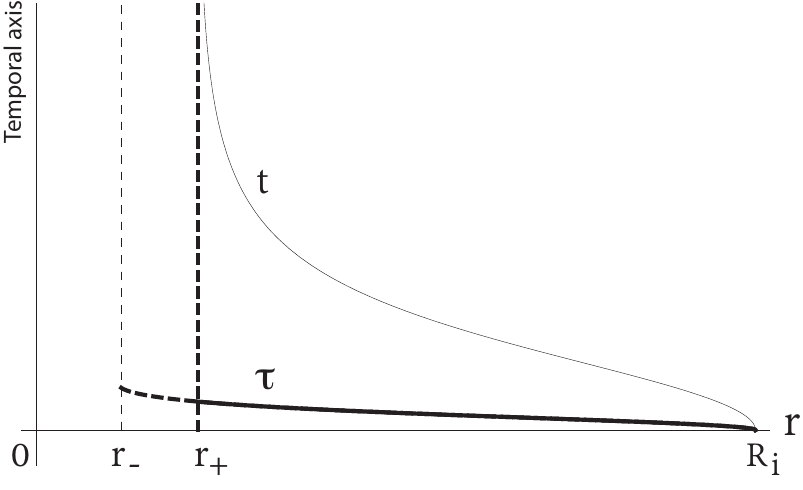}
	\end{center}
	\caption{The behavior of the proper $(\tau)$ and coordinate $(t)$ time for ingoing particles ($L>0$) as function of $r$, whit $M=2$, $a=1,9$, $L=4$, and $r_0=1$.}
	\label{f4.1}
\end{figure}
\newpage

The solution for $\phi(r)$ is given by
\begin{equation}
\phi(r)=-B_+\left[F_+(r)-F_+(R_i)\right]+B_-\left[F_-(r)-F_-(R_i)\right]
\end{equation}
with
\begin{equation}
B_{\pm}={L+r_{\pm}(r_0aE-2ML)\over 4(r_+-r_-)\sqrt{2ML^2-2aELr_0-r_0a^2}}~.
\label{tr6}
\end{equation}
\begin{figure}[!h]
	\begin{center}
		\includegraphics[width=60mm]{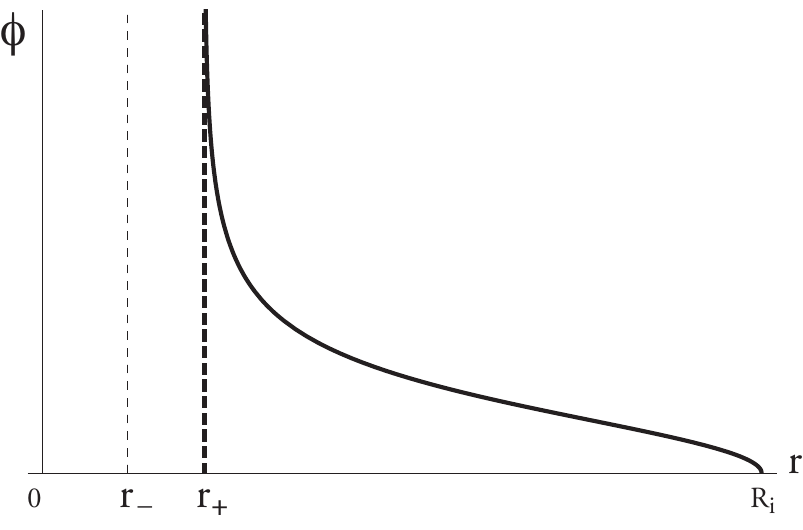}
        \includegraphics[width=40mm]{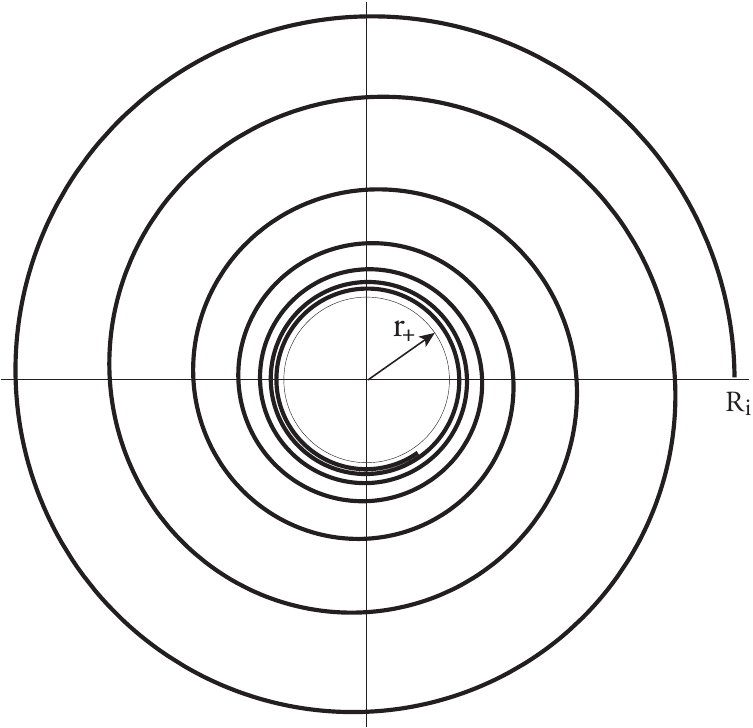}
	\end{center}
	\caption{The behavior of $\phi(r)$ for ingoing particles ($L>0$), with $M=2$, $a=1,9$, $L=4$, and $r_0=1$.}
	\label{f4.2}
\end{figure}
In Fig. \ref{f4.2} we plot the behavior of $\phi(r)$, and we can observe that a trajectory approaching the horizon will spiral round the black hole an infinite number of times such as was observed by Chandrasekhar for the geodesics in the Kerr spacetime \cite{chandra}.

\subsubsection{Radial Geodesics}

\begin{figure}[!h]
	\begin{center}
		\includegraphics[width=70mm]{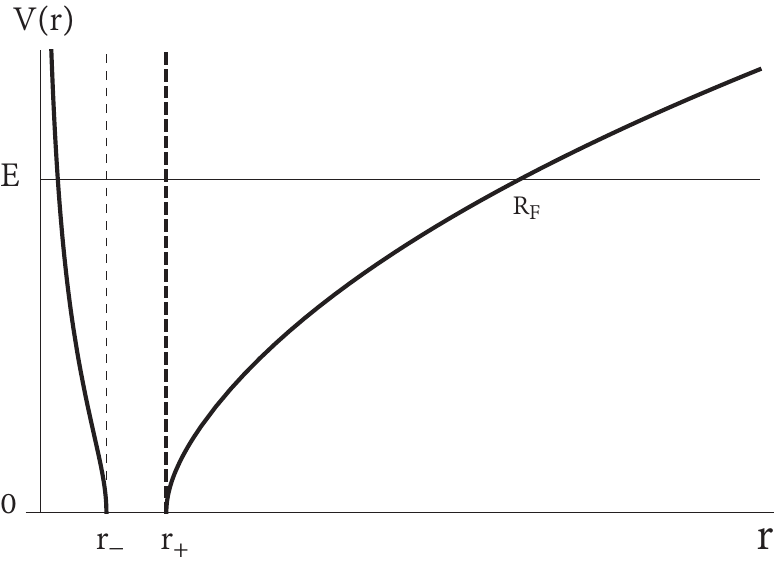}
	\end{center}
	\caption{The behavior of the effective potential for massive particles as a function of $r$, $r_0=1$, $M=2$, $L=4$, $\left| a\right| =1.9$, $R_F=10$ and $E=2.52$.}
	\label{RG}
\end{figure}

In this case $(L=0)$ is adequate to rewrite the effective potential in the
following form
\begin{equation}\label{tl82}
V(r)=\sqrt{{ r^2-2Mr+a^2\over r_0r} }~.
\end{equation}
Notice that the effective potential is the same for direct and retrograde geodesics.
The behaviors of the effective potential for massive particles is plotted in Fig \ref{RG}. The turning point, $R_{F}$, can be obtained from Eq. (\ref{ta2}) by imposing $L=0$, and is given by
\begin{equation}R_F= {r_0E^2+2M\over 2}\left( 1+\sqrt{1-{4a^2 \over (r_0E^2+2M)^2} }\right)~,
\label{ta22}
\end{equation}
It is worth to mention that  the radial motion allows only bounded orbits, similar to the trajectories with angular momentum. However, we can observe that the energy for particle without angular momentum must be positive $(E>0)$ independently if it is a direct o retrograde geodesics. In contrast to the trajectories with angular momentum where orbits with negative or null energy for retrograde trajectories are  possible. On the other hand, the solutions for $\phi(r)$, $\tau(r)$, and $t(r)$ have a similar behavior to the observed for particles with angular momentum due to the effective potential.

\subsection{Null geodesics in the equatorial plane}

In this section we analyze the motion
of photons, $m^2=0$, so the
effective potential is given by
\begin{equation}
V\left( r\right) =\frac{L}{r_0 r}\left( a+\sqrt{ r^2-2Mr+a^2}\right)~,
\label{t1}
\end{equation}
whose behavior is given in Fig. \ref{f4.1}  for photons with
$a>0$ (direct null geodesics) and $a<0$ (retrograde null geodesics), where $E_{\infty}=L/r_0$ corresponds to the energy of the photon when $r\rightarrow \infty$. $E_+= V(r_+, \left| a\right|)$ and $E_-= V(r_+, -\left| a\right|)$. Notice that retrograde null geodesics allow trajectories with negative energy with $E>E_-$. However, in this case, we can distinguish two zones, one of them allows bounded orbits $(E<E_{\infty})$ and the other one allows unbounded orbits $(E\ge E_{\infty})$.
\begin{figure}[!h]
	\begin{center}
		\includegraphics[width=70mm]{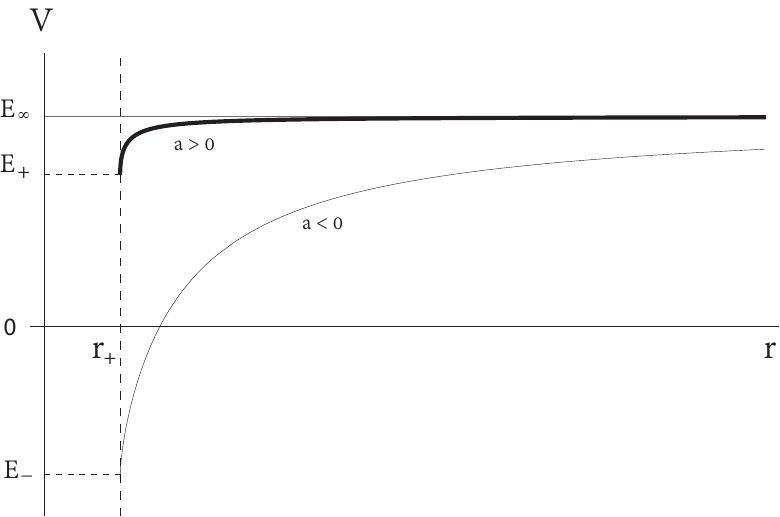}
	\end{center}
	\caption{The behavior of the effective potential for photons  ($L>0$) as a function of $r$, with $r_0=1$, $M=2$, $\left| a\right| =1.9$. }
	\label{f4.1}
\end{figure}
\newpage

\subsubsection{Direct null bounded trajectory}

\begin{figure}[!h]
	\begin{center}
		\includegraphics[width=70mm]{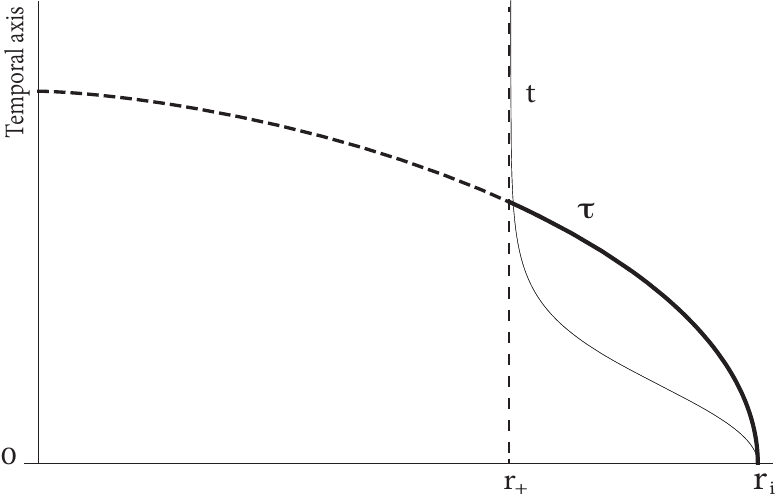}
	\end{center}
	\caption{ The behavior of the proper $(\tau)$ and coordinate $(t)$ time for ingoing photons ($L>0$) as function of $r$, with $r_0=1$, $M=2$, $r_i=4$, and $a=1.9$.}
	\label{f411}
\end{figure}

\begin{figure}[!h]
	\begin{center}
		\includegraphics[width=60mm]{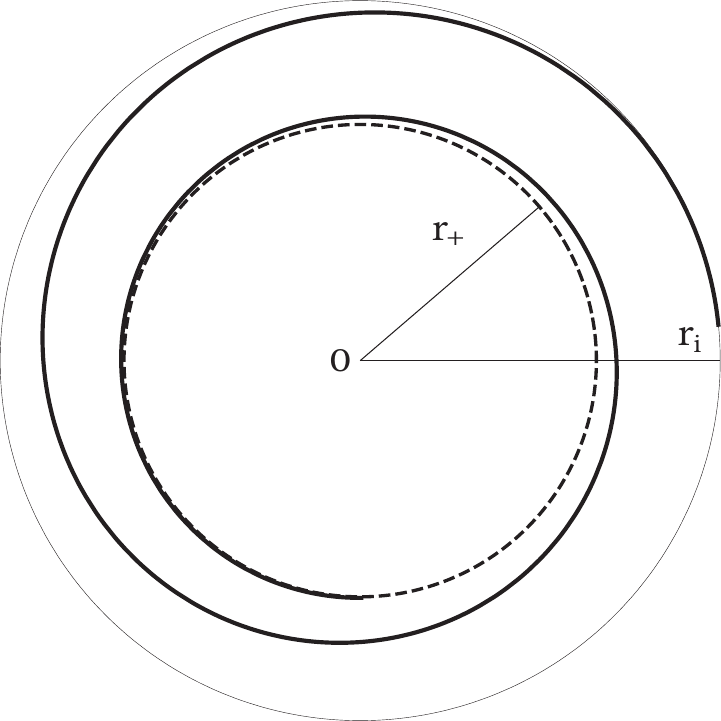}
	\end{center}
	\caption{The behavior of $\phi(r)$ for ingoing photons ($L>0$), with $r_0=1$, $M=2$, $r_i=4$, and $a=1.9$.}
	\label{f44}
\end{figure}

Now, we consider the direct null bounded trajectory, which is allowed for $E_+<E<E_{\infty}$, and the return point of the photon yields
\begin{equation}
r_i={2L(ML-ar_0E)\over L^2-r_0^2E^2}~.
\end{equation}
So, choosing the initial conditions for the photons as $r=r_i$
when $\phi=t=\tau=0$, Eq. (\ref{w.14}) and Eq. (\ref{w.12})  yields
\begin{equation}
\tau(r)=\frac{r_0}{\sqrt{L^2-E^2r_0^2}}\left[\sqrt{r_ir-r^2}
+r_i\tan^{-1}\sqrt{{r_i\over r}-1}\,\right]~,
\label{mr.3}
\end{equation}
\begin{eqnarray}
\notag t(r)&=&A_0\left[
2\tan^{-1}\sqrt{{r_i\over r}-1}\,+A_1\ln\left[{\sqrt{r(r_i-r_+)}+
	\sqrt{r_+(r_i-r)}\over\sqrt{r(r_i-r_+)}-\sqrt{r_+(r_i-r)}}\right]\right]+\\
&& A_0 A_2\ln\left[{\sqrt{r(r_i-r_-)}+
	\sqrt{r_-(r_i-r)}\over\sqrt{r(r_i-r_-)}-\sqrt{r_-(r_i-r)}}\right],
\label{mr.3}
\end{eqnarray}
where, the constants $A_0$, $A_1$ and $A_2$ are given by
\begin{equation}
A_0=\frac{r_0^2E}{\sqrt{L^2-E^2r_0^2}}~,
A_1={r_+(r_+-aL/r_0E)\over (r_+-r_-)\sqrt{r_+(r_i-r_+)}}~,
A_2={r_-(aL/r_0E-r_-)\over (r_+-r_-)\sqrt{r_-(r_i-r_-)}}~,
\end{equation}
and Eq. (\ref{w.13}) yields
\begin{eqnarray}
\notag \phi(r)&=&B_0\left[
2\tan^{-1}\sqrt{{r_i\over r}-1}\,+B_1\ln\left[{\sqrt{r(r_i-r_+)}+
	\sqrt{r_+(r_i-r)}\over\sqrt{r(r_i-r_+)}-\sqrt{r_+(r_i-r)}}\right]\right]+\\
&& B_0 B_2\ln\left[{\sqrt{r(r_i-r_-)}+
	\sqrt{r_-(r_i-r)}\over\sqrt{r(r_i-r_-)}-\sqrt{r_-(r_i-r)}}\right],
\label{mr.3}
\end{eqnarray}
and the  constants $B_0$, $B_1$, and $B_2$ are given by
\begin{equation}
B_0=\frac{L}{\sqrt{L^2-E^2r_0^2}}~,
B_1={r_+(r_+-2M+ar_0E/L)\over (r_+-r_-)\sqrt{r_+(r_i-r_+)}}~,
B_2={r_-(2M-ar_0E/L-r_-)\over (r_+-r_-)\sqrt{r_-(r_i-r_-)}}~.
\end{equation}
Then, in Fig. \ref{f411} we plot  the proper $(\tau)$ and coordinate time $(t)$ as function of $r$ for a photon falling from a finite distance, we can see that the photon falls towards the horizon in a finite proper time. The situation is very different if we consider the trajectory in the coordinate time, where t goes to infinity. Also, in Fig. \ref{f44} we plot the behavior of $\phi(r)$, and we can observe that a trajectory approaching the horizon will spiral round the black hole an infinite number of times.

\newpage

\subsubsection{Direct null unbounded trajectory}

Photons with energy $E_{\infty}\leq E$ have not a barrier of potential or turning point. So, the photons can plunge into the horizon or escape to the infinity. In this section, we consider the special case $E=E_{\infty}=V(\infty)=L/r_0$ by simplicity. Thus, choosing the following initial conditions for the photons: $\phi=t=\tau=0$, when $r=r_i$, the Eq. (\ref{w.14}), Eq. (\ref{w.12}) and  Eq. (\ref{w.13}) yield
\begin{equation}
\label{taup}
\tau(r)=\pm \frac{2k_0}{3}\left[r^{3/2}-r_i^{3/2}\,\right], \quad \text{and} \quad
k_0={r_0\over L\sqrt{2(M-a)}}~.
\end{equation}
\begin{equation}
t(r)=\pm{k_0L\over (r_{+}-r_{-})}
\left[(r_{+}-a)
[T_+(r)-T_+(r_i)]+
(a-r_{-})
[T_-(r)-T_-(r_i)]
\right]~,
\label{mr.5}
\end{equation}
and
\begin{equation}
\phi(r)=\pm{k_0L\over r_0(r_{+}-r_{-})}
\left[(r_{+}-2M+a)
[T_+(r)-T_+(r_i)]+
(2M-a-r_{-})
[T_-(r)-T_-(r_i)]
\right]~,
\label{mr.5}
\end{equation}
where
\begin{equation}
T_{\pm}(r)=2\sqrt{r}-2\sqrt{r_{\pm}}\tanh^{-1}
\sqrt{{r\over r_{\pm}}}~.
\end{equation}
The sign $\pm$ corresponds to trajectories that escape to the infinity or that plunge into the horizon. In section V, we will plot the unbounded trajectories for photons. Notice that from Eq. (\ref{taup}) in the proper frame  the photons arrive to the event horizon in a finite proper time.



\subsubsection{Radial Motion}

\begin{figure}[!h]
	\begin{center}
		\includegraphics[width=80mm]{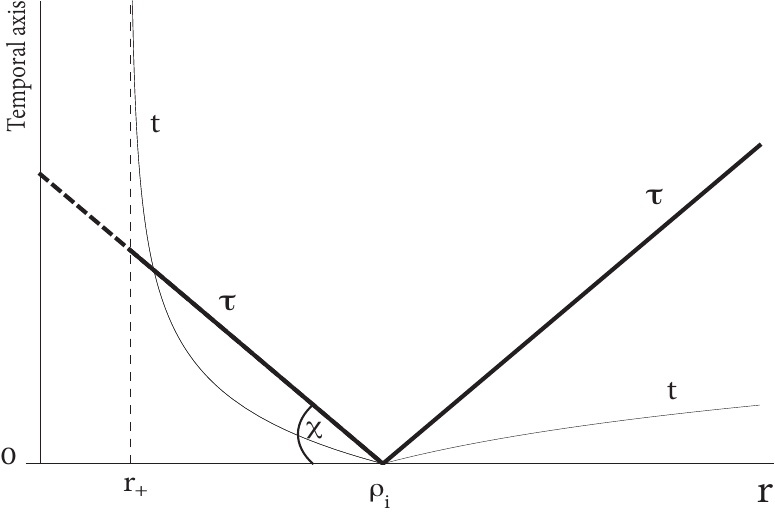}
	\end{center}
	\caption{The variation of the coordinate time $(t)$ and the proper time $(\tau)$ along an unbounded time-like radial geodesic described by a photon as test particle, starting at $\rho_i=10$ and falling toward the singularity or going towards infinity, for $M=2$, $a=1,9$, and $r_0= 1$. Thick line for the proper time and thin line for the coordinate time. The dashed part of the curve inside the horizon has no physical meaning.}
	\label{f444}
\end{figure}

\begin{figure}[!h]
	\begin{center}
	\includegraphics[width=60mm]{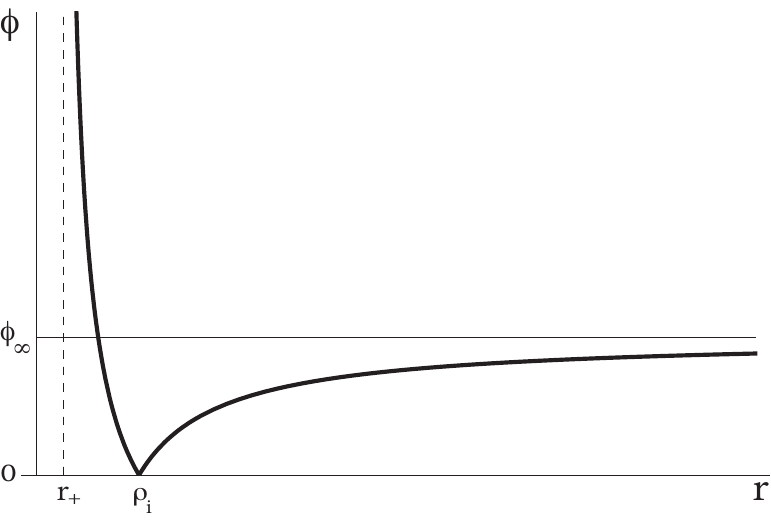}
	\includegraphics[width=70mm]{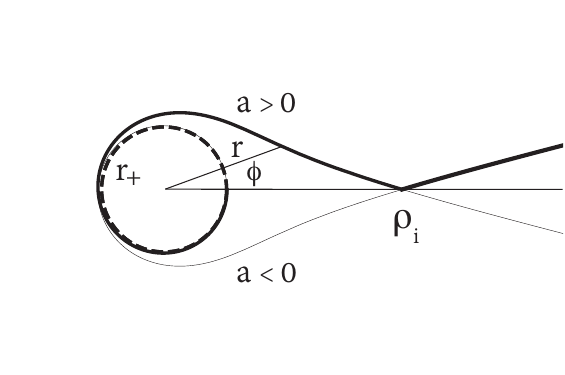}
	\end{center}
	\caption{The behavior for $\phi(r)$ for direct and retrograde trajectory starting at $\rho_i=10$ and falling toward the singularity or going towards infinity, for $M=2$, $a=1,9$, and $r_0=1$. }
	\label{f4444}
\end{figure}

Radial motion corresponds to a trajectory with null angular
momentum $L=0$, and photons are destined
to fall toward the event horizon. From Eq.
(\ref{tl8}) we can see that for radial photons we have
$V(r)=0$, so that eqs. (\ref{w.12}), (\ref{w.13}) and (\ref{w.14}) in the equatorial plane become
\begin{eqnarray}
\label{w.12r}
&&\dot{t}= { r_0rE \over r^2-2Mr+a^2}~,\\
\label{w.13r}
&&\dot{\phi}= {\,aE \over (r^2-2Mr+a^2)}~,\\
\label{w.14r}
&&\pm \dot{r}=E~,
\end{eqnarray}
where the sign ($-$) corresponds to photons  falling
into the event horizon and the sign ($+$) corresponds to photons that escape to infinity.
Choosing the initial conditions for the photons as $r=\rho_i$
when $\phi=t=\tau=0$, Eq. (\ref{w.14r}) yields
\begin{equation}
\tau(r)=\pm \frac{1}{E}(r-\rho_i)~,
\label{mr.3}
\end{equation}
where the sign $\pm$ has the same meaning given previously. Note that the above equation depends on the energy $E$ and it is valid for retrograde and direct geodesics. Also, for the negative sign the photons, in the proper frame, arrive to the event horizon in a finite proper time, see Fig. \ref{f444}. On the other hand, a straightforward integration of Eq. (\ref{w.12r}) and (\ref{w.13r}) leads to
\begin{equation}
t(r)=\pm {r_0\over r_{+}-r_{-}}\left[r_{+}
\ln \left|\frac{r-r_{+}}{\rho_i-r_{+}}\right|-
r_{-}
\ln \left|\frac{r-r_{-}}{\rho_i-r_{-}}\right|
\right]~,
\label{mr.5}
\end{equation}
and
\begin{equation}
\phi(r)=\pm {a\over r_{+}-r_{-}}\left[
\ln \left|\frac{r-r_{+}}{\rho_i-r_{+}}\right|-
\ln \left|\frac{r-r_{-}}{\rho_i-r_{-}}\right|
\right]~.
\label{mr.5b}
\end{equation}
Note that the solution for the coordinate time $t$ does not depend on the energy of the photon and the expression is valid for retrograde and direct geodesics. While  the solution for $\phi$, see Fig. \ref{f4444}, also, it does not depend on the energy of the photon but it depends on the angular momentum parameter $a$. Moreover, for direct geodesic when $r \rightarrow \infty$
\begin{equation}
\phi_{\infty} = \frac{a}{r_+-r_-}\ln  \left|\frac{\rho_i -r_{-}}{\rho_i-r_{+}}\right|~.
\end{equation}

It is worth to mention that another remarkable feature of the motion of photons is the existence of circular orbits of any radius. This occurs in the extremal case $(M=a)$  and for $L=Er_0$.

\section{The CM energy of two colliding particles}
\label{CMS}

In this section we study the CM energy of two colliding massive particles in the equatorial plane $\theta = \pi/2$,
the 4-velocity of the test particles is given by $u=(\dot{t},\dot{r},0,\dot{\phi})$, where
\begin{eqnarray}
\label{tdot} \dot{t} &=& \frac{r_0 r E-aL}{r_0 r f(r)}~, \\
\label{rdot} \dot{r}^2 &=& \left(  E-\frac{aL}{r_0 r}\right)^2-f(r) \left( 1+\frac{L^2}{r_0 r} \right)~, \\
\label{pdot}\dot{\phi} &=& \frac{a E r_0 r+L r_0 r f(r) - a^2 L}{r_0^2 r^2 f(r)}~.
\end{eqnarray}
We will assume that $\dot{t}>0$ for all $r>r_+$ so that the motion is forward in time outside the horizon, so:
\begin{equation}
r_0 r E-aL>0, \,\, \text{for all}\,\, r>r_+~.
\end{equation}
Now, we use the 4-velocity components to obtain the CM energy of the collision of two particles in the linear dilaton black hole, and we consider that the particles have energies $E_1$ and $E_2$ and angular momentum $L_1$ and $L_2$, respectively. Thus, using the relation $E_{CM}=\sqrt{2} m_0 \sqrt{1-g_{\mu \nu}u_1^{\mu}u_2^{\nu}}$, where $m_0$ denotes the rest mass of the particles, and $u_1$ and $u_2$ denotes the 4-velocities of the particles, we obtain
\begin{equation}\label{CM}
\frac{E_{CM}^2}{2 m_0^{2}}= \frac{r_0 r f(r) (r_0 r -L_1 L_2 )+(K_1 K_2-H_1 H_2)}{r_0^{2} r^{2} f(r)}~,
\end{equation}
where
\begin{eqnarray}
\notag K_i  &=& E_i r_0 r-a L_i~, \\
H_i &=& \sqrt{(E_i r_0 r-aL_i)^2-r_0 r f(r) (r_0 r+L_i^2)}~,
\end{eqnarray}
whit $i=1,2$. When the particles arrive to the horizon $r=r_{+}$, $f(r) \rightarrow 0$, $H_1 \rightarrow \sqrt{K_1^2}$ and $H_2 \rightarrow \sqrt{K_2^2}$, therefore:
\begin{equation}
\frac{E_{CM}^2}{2 m_0^{2}}(r \rightarrow r_{+})=\frac{1}{r_0^2 r^2 f(r)} (K_1 K_2 -\sqrt{K_1^2} \sqrt{K_2^2})~.
\end{equation}
Notice that for $K_1K_2<0$ the $E_{CM}$ on the horizon ($f(r)=0$) is negative infinite which is not a physical solution. However, when $K_1K_2 >0$, the numerator of  this expression will be zero and the value of $E_{CM}$ will be undetermined. In order to obtain the limiting value we use the L'Hopital rule, and obtain
\begin{equation}
\frac{E_{CM}^2}{2 m_0}=\frac{r_0 r_+ (K_1(r_+)+K_2(r_+))^2 +(K_1(r_+) L_2-K_2(r_+)L_1)^2}{2r_0 r_{+} K_1(r_+) K_2(r_+)}~.
\end{equation}
Note that the numerator of the above expression is finite at the horizon and when $K_i(r_+)=0$ the CM energy of two colliding particles on the horizon could be arbitrary high $E_{CM} \big | _{K_i=0} \rightarrow \infty$. From $K_i(r_+)=0$ we obtain the critical angular momentum:
\begin{equation}
L_{ci}= \frac{r_0 r_+ E_i}{a}~,  \,\,\,\,\, i=1,2~.
\end{equation}
Besides, when $K_1(r_+)$ and $K_2(r_+)$ are both zero, then $E_{CM}$ is finite. In this case $H_1(r_+)=H_2(r_+)=0$ and
\begin{equation}
\frac{E_{CM}^2}{2 m_0}=1-\frac{L_1L_2}{r_0 r_+}~.
\end{equation}
Therefore, in order to obtain infinite CM energy only one of the colliding particles must have the critical angular momentum.

By a similar analysis, it is possible to evaluate the $E_{CM}$  on the inner horizon, finding that this is also infinite as long as  one of the two particles has the following  critical angular momentum:
\begin{equation}\label{critical}
L_{ci}= \frac{r_0 r_- E_i}{a}~,  \,\,\,\,\, i=1,2~.
\end{equation}
Now, in order to see if a particle with critical angular momentum could reach the horizon, we shall analyze the radial motion of the particle with critical angular momentum and energy $E$, and we shall find the region where it can exist. The radial equation can be written as
\begin{equation}
\dot{r}^2=R(r)~,
\end{equation}
where
\begin{eqnarray}
\notag R(r)&=&\left(  E-\frac{aL}{r_0 r}\right)^2-f(r) \left( 1+\frac{L^2}{r_0 r} \right) \\
\label{aa} &=& -\frac{r}{r_0}+\frac{2ML^2-r_0 a^2-2 r_0 a E L}{r_0^2 r}+\frac{-L^2+2 r_0 M +r_0^2 E^2}{r_0^2}~.
\end{eqnarray}
Note that the particle can exist only in regions where $R(r) \geq 0$. Also, note that at the horizon $R(r_+) \geq 0$ and $R(r_+) = 0$ for a particle with the critical momentum $L_c=r_0 r_+ E/a$. Therefore, from expression (\ref{aa}) we find that if $2ML^2-r_0 a^2-2 r_0 a E L >0$ the radial function has the following asymptotic behaviors $R(r \rightarrow \infty) \rightarrow -\infty$ and $R(r \rightarrow 0) \rightarrow \infty$, so the function $R(r)$ has only one positive root, say $r_\ast^+$; thus, the particle can exist only in the region $0<r<r_\ast$. On the other hand, if $2ML^2-r_0 a^2-2 r_0 a E L <0$, we find $R(r \rightarrow \infty) \rightarrow -\infty$ and $R(r \rightarrow 0) \rightarrow -\infty$, so the function $R(r)$ has two positive roots, and the particle can exist in the region between the roots. The roots of $R(r)$ are given by:
\begin{equation}
r_\ast^\pm=\frac{E^2 r_0^2+2M r_0-L^2 \pm \sqrt{8 r_0 M L^2 -4 a^2 r_0^2 -8 a r_0^2 L E +(E^2 r_0^2+2M r_0-L^2)^2 }}{2r_0}~.
\end{equation}

For a particle with the critical angular momentum, the radial function is given by:
\begin{equation}\label{POTC}
R^{c}=R(r)|_{L=L_c}=\frac{r-r_+}{r} \left( -\frac{r-r_-}{r_0} +\frac{E^2 \left( a^2 -r_+^2 \right)}{a^2} \right)~,
\end{equation}
and it vanishes on the event horizon, to simplify the above equation we have used $r_+ r_-= a^2$.


The roots of the radial function $R(r)|_{L=L_c}$ are given by:
\begin{eqnarray}
r_1 &=& r_+ \\
\notag r_2 &=& r_- + \frac{r_0 E^2 (a^2-r_+^2)}{a^2} ~.
\end{eqnarray}

Particles with critical angular momentum can reach the event horizon if
\begin{equation}
\frac{dR^{c}}{dr}\Big|_{r=r_{+}}>0\, .
\end{equation}
For the linear dilaton black hole we obtain
\begin{equation}
\frac{dR^{c}}{dr}\Big|_{r=r_{+}}=-\frac{\left(r_+-r_-\right)  \left( 1+r_0 r_+ E^2 / a^2  \right)}{r_0 r_+}<0~,
\end{equation}
therefore, if particles with critical angular momentum exist outside the black hole, they cannot reach the event horizon. In Fig. \ref{f5} we plot the behavior of $R^c$ as a function of $r$ for the non-extremal linear dilaton black hole. Notice that if $\frac{d^{2}R^{c}}{dr^{2}} \Big|_{r=r_{+}}=2 \left( \frac{E^2}{a^2} -\frac{E^2}{r_+^2} -\frac{r_-}{r_0 r_+^2}\right)<0$, then $r_2$ is positive and smaller than $r_-$ and if $\frac{d^{2}R^{c}}{dr^{2}} \Big|_{r=r_{+}}=2 \left( \frac{E^2}{a^2} -\frac{E^2}{r_+^2} -\frac{r_-}{r_0 r_+^2}\right)>0$,
then $r_2$ is negative; therefore, particles with critical angular momentum cannot exist outside the event horizon.

 On the other hand, we notice that the particle with the critical angular momentum can exist inside the event horizon $r_{+}$. This can be shown by replacing the critical angular momentum Eq. (\ref{critical}) in the square of the radial component of the 4-velocity Eq. (\ref{rdot}), obtaining the analog of Eq. (\ref{POTC}), whose derivative evaluated on the internal horizon $r_-$ is positive $\frac{dR^{c}}{dr}\vert_{r=r_-}>0$; therefore, the particle with critical angular momentum can reach the inner  horizon and the CM energy can be arbitrarily high, with the BSW process  being possible on the inner horizon. This behavior is similar to the rotating BTZ black hole \cite{Lake:2010bq, Yang:2012we, Tsukamoto:2017rrl}. In Table \ref{CM} we show the behavior of the CM energy on the inner horizon as a function of the angular momentum $L_1$ of a particle, we observe that when $L_1$ approaches to the critical angular momentum $L_c=\frac{r_0 r_- E}{a}$, the CM energy increases and tends to infinity when $L_1 \rightarrow L_c$. Also, in table \ref{LCr0} we show the critical angular momentum $L_c=\frac{E_1 r_0 r_-}{a}$ and $e^{-2\phi}(r)$ for different values of $r_0$. Notice that in the equatorial plane the string coupling depends on $r_0$ and it does not depend on $a$. We also  observe that when the coupling $r_0$ is increasing, the  string coupling becomes stronger and this leads to a higher value of the critical angular momentum. Thus, a strong string coupling  requires the particle to have a high angular momentum in the collision process in order  the rotating linear dilaton black hole to act as a particle  accelerator.

\begin{table}[ht]
\caption{The behavior of $\frac{E_{CM}^2}{2m_0}$ as a function of $L_1$ for $a=1.9$, $M=2$, $r_0=1$, $E_1$=1, $E_2=2$ and $L_2=3$. The critical angular momentum is given by $L_c=\frac{E_1 r_0 r_-}{a}= 0.723947$ . }
\label{CM}\centering
\begin{tabular}{ | c | c | c | c |}
\hline
$L_1$ & $\frac{E_{CM}^2}{2m_0}$ & $L_1$ & $\frac{E_{CM}^2}{2m_0}$ \\ \hline
$5.0$ & $3.97$ & $0.73$ & $177.32$ \\  \hline
$4.0$ & $3.23$ & $0.729$ & $212.36$ \\  \hline
$3.0$ & $2.56$ & $0.728$ & $264.71$ \\  \hline
$2.0$ & $2.12$ & $0.727$ & $351.34$ \\  \hline
$1.0$ & $4.34$ & $0.726$ & $522.40$ \\  \hline
$0.9$ & $6.47$ & $0.725$ & $1018.50$ \\  \hline
$0.8$ & $14.39$ & $0.724$ & $20404.30$ \\  \hline
$0.75$ & $41.40$ & $0.72395$ & $424247.95$ \\  \hline
$0.74$ & $67.02$ & $0.723948$ & $2.03 \cdot 10^6$ \\  \hline
\end{tabular}%
\end{table}

\begin{table}[ht]
\caption{Critical angular momentum $L_c=\frac{E_1 r_0 r_-}{a}$ and $e^{-2\phi}(r)$ for different values of $r_0$, for $a=1.9$, $M=2$, and $E_1$=1. }
\label{LCr0}\centering
\begin{tabular}{ | c | c | c |}
\hline
$r_0$ & $e^{-2\phi}(r)$ & $L_c$ \\ \hline
$0.5$ & $\frac{1}{2r}$ & $0.263$  \\  \hline
$1.0$ & $\frac{1}{r}$ & $0.526$  \\  \hline
$1.5$ & $\frac{3}{2r}$ & $0.789$ \\  \hline
$2.0$ & $2r$ & $1.053$  \\  \hline
\end{tabular}%
\end{table}

 On the other hand, in the extremal case, the function $R^c(r)$ for a particle with critical angular momentum reads:
\begin{equation}
R^c(r)=-\frac{(r-M)^2}{r_0 r}~.
\end{equation}
So, the function $R^c(r)$ is null for $r=r_+=M$, which is the degenerate horizon of the extremal black hole. Notice that the particle with critical angular momentum cannot exist outside the event horizon; however, it can exist on the degenerate horizon, which is similar to the behavior for the extremal BTZ black hole \cite{Lake:2010bq, Yang:2012we, Tsukamoto:2017rrl}.

\begin{figure}[!h]
\begin{center}
\includegraphics[width=80mm]{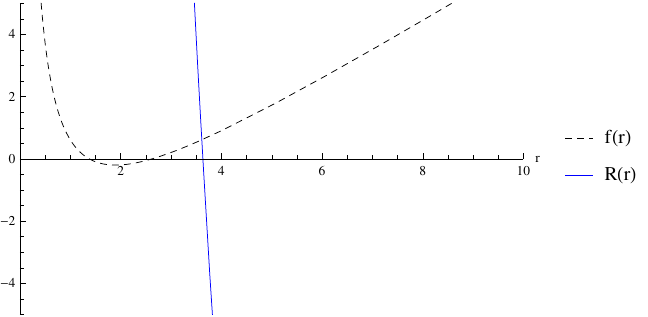}
\end{center}
\caption{The behavior of $f(r)$ and $R(r)$ for $a=1.9$, $M=2$, $r_0=1$, $E=1$ and $L=10$.}
\label{f1}
\end{figure}

\begin{figure}[!h]
\begin{center}
\includegraphics[width=80mm]{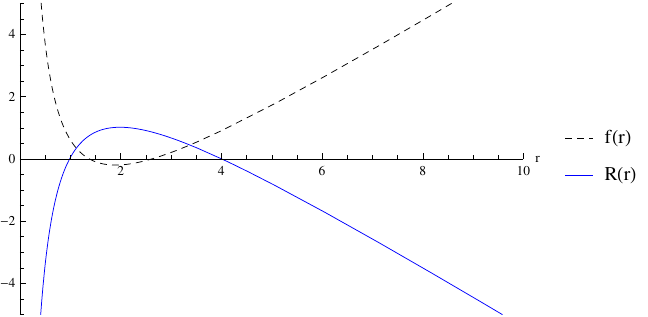}
\end{center}
\caption{The behavior of $f(r)$ and $R(r)$ for $a=1.9$, $M=2$, $r_0=1$, $E=1$ and $L=0.1$.}
\label{f2}
\end{figure}

\begin{figure}[!h]
\begin{center}
\includegraphics[width=80mm]{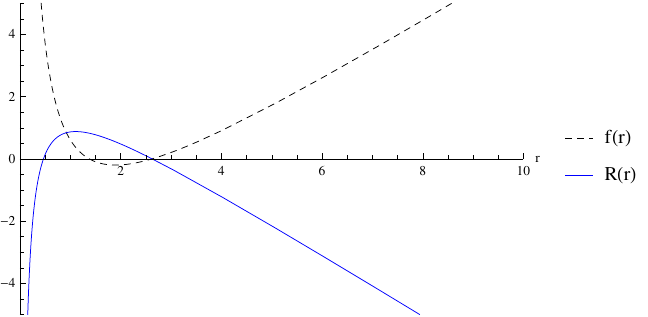}
\end{center}
\caption{The behavior of $f(r)$ and $R(r)$ for $a=1.9$, $M=2$, $r_0=1$, $E=1$ and $L=L_c$.}
\label{f5}
\end{figure}

\begin{figure}[!h]
\begin{center}
\includegraphics[width=80mm]{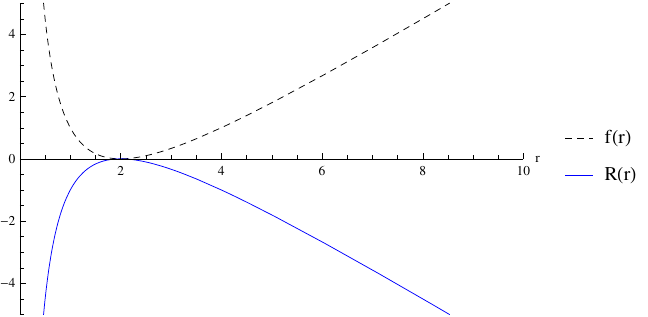}
\end{center}
\caption{The behavior of $f(r)$ and $R(r)$ for $a=2$, $M=2$, $r_0=1$, $E=1$ and $L=L_c$.}
\label{f6}
\end{figure}

Now, in order to illustrate these behaviors, we show some graphics. In Fig. \ref{f1} we have plotted the behavior of $f(r)$ and $R(r)$ for $a=1.9$, $M=2$, $r_0=1$, $E=1$ and $L=10$, and in Fig. \ref{f2} we have plotted the behavior of $f(r)$ and $R(r)$ for $a=1.9$, $M=2$, $r_0=1$, $E=1$ and $L=0.1$. In Figs. \ref{f5} and \ref{f6} we show the behavior of $R^c(r)$ for a particle with critical angular momentum in the non-extremal and in the extremal case, respectively. One root of $R^c(r)$ coincide with the horizon in the non-extremal case. This means that the particle with the critical angular momentum can exist inside the outer horizon and the particle collision on the inner horizon could produce arbitrarily high CM energy. In the extremal case the roots of $R^c(r)$ coincide with the degenerate horizon, so, particles with critical angular momentum can only exist on the degenerate horizon.

\begin{figure}[!h]
\begin{center}
\includegraphics[width=80mm]{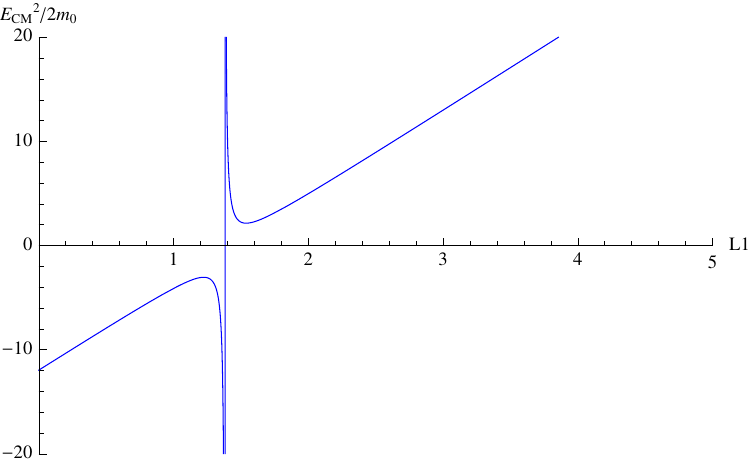}
\end{center}
\caption{The behavior of $E_{CM}$ as a function of $L_1$, for $a=1.9$, $M=2$, $r_0=1$, $E_1=1$, $E_2=2$ and $L_2=3$.}
\label{f3}
\end{figure}

\begin{figure}[!h]
\begin{center}
\includegraphics[width=80mm]{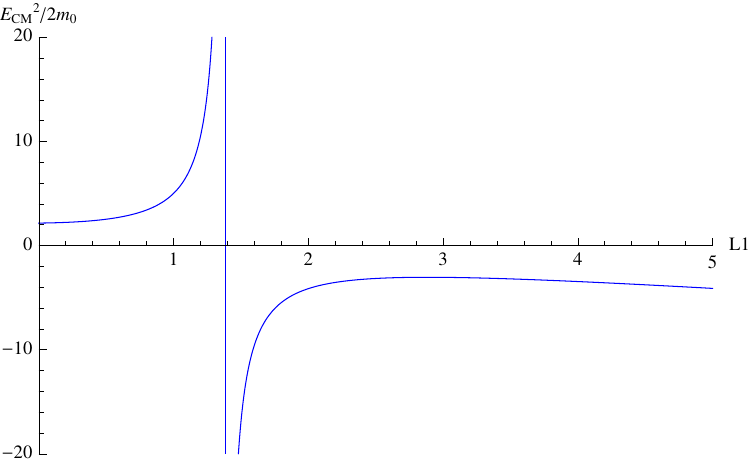}
\end{center}
\caption{The behavior of $E_{CM}$ as a function of $L_1$, for $a=1.9$, $M=2$, $r_0=1$, $E_1=1$, $E_2=2$ and $L_2=0.5$.}
\label{f4}
\end{figure}

In Fig. \ref{f3} we have plotted the behavior of $E_{CM}$ as a function of $L_1$, with $a=1.9$, $M=2$, $r_0=1$, $E_1=1$, $E_2=2$ and $L_3=3$. We see that in order to get infinite CM energy, particle 1 reach the  critical angular momentum as shown in Figs. \ref{f3} and \ref{f4}.

\newpage

\section{The general equations of geodesic motion}
\label{G4D}
In this section we solve analytically the geodesic equations of motion of the test particle for the radial, temporal and orbital motion.
\subsection{Analysis of the angular motion ($\theta$-motion)}
In order to study the $\theta$-motion, we consider the equation of motion (\ref{angular}), which can be rewritten as  $d \theta/d
\gamma=\sqrt{\Theta}$, where the coordinate $\theta$ is a polar angle that can take only positive values. So
\begin{equation}
\Theta=k^2-\frac{L^2}{\sin ^2 \theta} \geq 0~,
\end{equation}
where the separability constant is non null or negative. Now, through the change of variables $\xi=\cos\theta$, Eq. (\ref{angular}) yields
\begin{equation}\label{eqxi}
\frac{d \xi}{d \gamma}=-\sqrt{\Theta_{\xi}}~,
\end{equation}
being $\Theta_{\xi}=k^2(1-\xi^2)-L^2$. The roots of the function $\Theta_{\xi}$ are given by
\begin{equation}
\theta_1=\arccos \left( \sqrt{1-\frac{L^2}{k^2}} \right), \,\,\, \theta_2=\arccos \left( -\sqrt{1-\frac{L^2}{k^2}}\right)~,
\end{equation}
which define the cone's angles that confine the movement of the particle.  Integrating \eqref{eqxi} we get
\begin{equation}
\gamma(\xi)=- \int \frac{d \xi}{\sqrt{k^2(1-\xi^2)-L^2}}~.
\end{equation}
The solution, in terms of the variable $\theta$ is:
\begin{equation}
\gamma (\theta)= \frac{1}{\sqrt{k^2}}\arccos \left( \frac{\cos \theta}{\sqrt{1-\frac{L^2}{k^2}}} \right)~,
\end{equation}
where we have used that $\gamma_{0}=0$ for $\theta_{0}=\theta_{1}$.
Also, the above equation can be inverted, which yields
\begin{equation}
\theta (\gamma)= \arccos \left( \sqrt{1-\frac{L^2}{k^2}} \cos (\sqrt{k^2} \gamma) \right)~,
\end{equation}
for $k^2>L^2$.

\subsection{Analysis of the radial motion ($r$-motion) }

Now, we consider the motion of the particle with respect to the $r$-coordinate. We will focus in Eq. (\ref{eqradial}) in order to obtain the velocity of the particle $dr/dt$. The condition of  turning point $(\frac{dr}{dt})_{r=r_t}=0$ allows us to define an effective potential
\begin{equation}
\left( E-\frac{aL}{r_0 r}\right)^2-f(r) \left( m^2+ \frac{k^2}{r_0 r}\right)=(E-V_-)(E-V_+)~,
\end{equation}
where we can recognize the effective potential for the particle
with mass $m$ as
\begin{equation}
V_{\pm}=\frac{aL}{r_0 r} \pm \sqrt{f(r)\left( m^2+\frac{k^2}{r_0 r} \right)}~.
\end{equation}
The behavior of the effective potential is
shown in Fig. \ref{fig.1}.
In the following sections, we will consider the motion of particles $m \neq 0$ and the motion of photons $m =0$ separately.

\subsubsection{Radial motion of particles}

Now, Eq. \eqref{eqradial} can be written as:
\begin{equation}
\left( \frac{dr}{d \gamma} \right)^2=\,P(r)~,
\end{equation}
where $P(r)=-a_3\,r^3+a_2 r^2+a_1 r+a_0$, with
\begin{equation}
a_{0}=-a^2 (k^2-L^2), \,\,\, a_{1}=2k^2 M-2aEL r_0 -a^2 m^2r_0, \,\,\, a_{2} =-k^2+2m^2 Mr_0+E^2 r_0^2, \,\,\, a_{3}=m^2r_0>0~.
\end{equation}
Integrating the above equation, where $r_1$ corresponds to the starting point, we obtain
\begin{equation}
\int _{\gamma_i=0}^{\gamma} d\gamma = {1\over \sqrt{a_3}} \int _{r_1}^r \frac{-dr}{\sqrt{P(r)/a_3}}~,
\end{equation}
with the change of variable $r=-4\,x+\frac{a_{2}}{3a_{3}}$, the integral transform to
\begin{equation}
\gamma(r)=
\frac{1}{\sqrt{a_3}}\int_{x_1}^{x} \frac{dx}{\sqrt{4x^3-g_{2}x-g_{3}}} ~,
\end{equation}
where $r_{1}=-4\,x_{1}+\frac{a_{2}}{3a_{3}}$. The invariants are:
\begin{equation}
g_{2}=\frac{1}{a_{3}}\left( \frac{a_{2}^2}{3a_{3}}+a_{1} \right), \,\,\, g_{3}=-\frac{1}{16a_{3}} \left( a_{0}+\frac{2a_{2}^3}{27 a_{3}^2}+\frac{a_{1} a_{2}}{3a_{3}} \right)~,
\end{equation}
\begin{equation}
\gamma=\frac{1}{\sqrt{a_{3}}} \left(\wp^{-1} (x; g_{2},g_{3}) -\varphi_1\right)~,
\end{equation}
being $ \varphi_1=\wp^{-1} (x_{1}; g_{2},g_{3})$. Inverting this expression, we obtain:
\begin{equation}
r(\gamma)=\frac{a_{2}}{3a_{3}}-4 \wp \left(  \sqrt{a_{3}} \,\gamma+  \varphi_1; g_{2},g_{3} \right)~.
\end{equation}

\subsubsection{Bounded radial motion of photons}

For photons falling from $\rho_1$,
\begin{equation}
\int _{0}^{\gamma} d\gamma = \frac{1}{\sqrt{b_{2}}}  \int _{\rho_1}^r \frac{-dr}{\sqrt{P_2(r)/b_2}}~,
\end{equation}
where $P_2(r)=-b_2 r^2+b_1 r-b_0$, being
\begin{eqnarray}
\notag b_{0}&=&a^2 (k^2-L^2), \,\,\, b_{1}=2k^2 M-2aEL r_0, \\
&& b_{2} =k^2-E^2 r_0^2,\, b_4=b_1^2-4b_0b_2~,
\end{eqnarray}
\begin{equation}
\gamma=\frac{1}{\sqrt{b_{2}}} \left(\arcsin \left({b_1-2b_2\,r\over \sqrt{b_4} }\right) -\varphi_2\right)~,
\end{equation}
with $\varphi_2=\arcsin \left({b_1-2b_2\,\rho_1\over \sqrt{b_4} }\right) $.
Inverting this expression, we obtain:
\begin{equation}
r(\gamma)=\frac{b_{1}}{2b_{2}}-\frac{\sqrt{b_{4}}}{2b_{2}}\sin \left(  \sqrt{b_{2}} \,\gamma+  \varphi_2\right)~.
\end{equation}

\subsubsection{Unbounded radial motion of photons}

In this case we will consider $E=k/r_0$ and the solution is given by
\begin{equation}
r(\gamma)= \frac{b_0}{c_1}+\frac{1}{c_1}\left(\varphi_3 \pm \frac{c_1}{2}\gamma\right)^2~,
\end{equation}
where $c_1=2k(Mk-aL)$, $\varphi_3=\sqrt{c_1\rho_1-b_0}$, and the sign $(+)$ corresponds to photons that escape to infinity and the sign $(-)$, for photons that fall into the horizon.

\subsection{Analysis of the angular motion ($\varphi$-motion)}

From Eq. (\ref{varphi}) is possible to distinguish an integral in the radial part and the second one in the angular $\theta$ part. So the solution can be written as $\varphi=\varphi^{m}(r)+\varphi(\theta)$. The solution of the angular integral is given by
\begin{equation}
\varphi (\theta)= \frac{\pi}{2}+\frac{1}{2}
\arcsin\left[\frac{\cos^2\theta_1-\cos\theta}{\cos\theta_1(1-\cos\theta)}\right]
-\frac{1}{2}\arcsin\left[\frac{\cos^2\theta_1+\cos\theta}{\cos\theta_1(1+\cos\theta)}\right]~,
\end{equation}
and for particles the radial integral is given by
\begin{equation}
\varphi^{m=1}(r)=\frac{a\sqrt{r_0}E}{4m(r_+-r_-)}\left[(r_+-\frac{aL}{r_0E})(\bar{F}_+(u_1)-\bar{F}_+(u))-(r_--\frac{aL}{r_0E})(\bar{F}_-(u_1)-\bar{F}_-(u))\right]~,
\end{equation}
where
\begin{equation}
\bar{F}_{\pm}(r)=\frac{2\zeta(W_{\pm})\wp^{-1}[u(r)]}{\wp^{'}(W_{\pm})}
+\frac{1}{\wp^{'}(W_{\pm})}\ln\left| \frac{\sigma[\wp^{-1}[u(r)]-W_{\pm}]}
{\sigma[\wp^{-1}[u(r)]+W_{\pm}]}\right|
~,\label{tr444}\end{equation}
\begin{equation}
u(r)=\frac{a_2}{12 a_3}-\frac{r}{4}~,
\end{equation}
$u_1=u(r_1)$ and
\begin{equation}
W_\pm=\wp^{-1}\left[ \frac{a_2}{12 a_3}-\frac{r_\pm}{4}\right]~.
\end{equation}
On the other hand, for bounded photons the solution is:
\begin{equation}
\varphi^{m=0}(r)=\frac{ar_0E}{\sqrt{b_2}(r_+-r_-)}
\left[(r_+-\frac{aL}{r_0E})(\bar{G}_+(r)-\bar{G}_+(\rho_1))-
(r_--\frac{aL}{r_0E})(\bar{G}_-(r)-\bar{G}_-(\rho_1))\right]~,
\end{equation}
being
\begin{equation}
\bar{G}_{\pm}(r)=
+\frac{1}{\sqrt{\gamma_{\pm}}}
\cosh^{-1}\left| \frac{2\gamma_{\pm}+\delta_{\pm}(r-r_{\pm})}
{(r-r_{\pm})\sqrt{4\gamma_{\pm}+\delta_{\pm}^{2}}}\right| ~,
~,\label{tr444}\end{equation}
and
$$
\gamma_{\pm}={b_1\over b_2}r_{\pm}-{b_0\over b_2}-r_{\pm}^2~,
$$
$$
\delta_{\pm}={b_0\over b_2}-2r_{\pm}~.
$$
While that for unbounded photons the solution is
\begin{equation}
\varphi^{(0)}(r)=\frac{ar_0E}{(r_+-r_-)}
\left[(r_+-\frac{aL}{r_0E})(\bar{J}_+(r)-\bar{J}_+(\rho_1))-
(r_--\frac{aL}{r_0E})(\bar{J}_-(r)-\bar{J}_-(\rho_1))\right]~,
\end{equation}
where
\begin{equation}
\bar{J}_{\pm}(r)=
\frac{2}{\sqrt{b_0-c_1r_{\pm}}}
\tan^{-1}\left| \frac{\sqrt{c_1r-b_0}}
{\sqrt{b_0-c_1r_{\pm}}}\right|
~.\label{tr444}\end{equation}

\subsection{Analysis of the time motion ($t$-motion) }

The solution for $t$-motion of particles is given by the following expression
  \begin{equation}
 t^{(m=1)}(r)=\frac{a\sqrt{r_0}E}{r_-}\left[
 \wp^{-1} (u)- \wp^{-1} (u_A)
 +\frac{r_+}{4}
 [\bar{F}_+(u_A)-\bar{F}_+(u)]
 \right]~,
 \end{equation}
where we have considered by simplicity that the energy of the particle is $E_A=V(r_-)=V(R_A)=aL/r_0r_-$, being $R_A$ the turning point or Apoastro, and  $u_A=u(R_A)$. On the other hand,  the solution for $t$-motion of bounded photons yields
\begin{equation}
t^{(m=0)}(r)=\frac{ar_0E}{r_-\sqrt{b_{2}}}\left[
\arcsin \left({b_1-2b_2\,r\over \sqrt{b_4} }\right) - \arcsin \left({b_1-2b_2\,\rho_A\over \sqrt{b_4} }\right)
+\frac{r_+}{\sqrt{\gamma_{+}}}
[\bar{G}_+(r)-\bar{G}_+(\rho_A)]
\right]~,
\end{equation}
where we have considered by simplicity that the energy of the photon is $E_A=V(r_-)=V(\rho_A)=aL/r_0r_-$, being $\rho_A$ the turning point or Apoastro. While that for unbounded photons the solution of $t$-motion is given by
\begin{equation}
t^{(0)}(r)=\pm\frac{ar_0E}{r_-}\left[
\frac{2}{c_1}
[\sqrt{c_1\,r-b_0}-\sqrt{c_1\rho_A-b_0}]
+r_+
[\bar{J}_+(r)-\bar{J}_+(\rho_A)]
\right]~,
\end{equation}
where we have considered by simplicity that the energy of the photon is $E_A=V(r_-)=V(\rho_A)=aL/r_0r_-$,  and $k=aL/r_-$. Being, $\rho_A$ the starting point of the trajectory.

\begin{figure}[!h]
\begin{center}
\includegraphics[width=60mm]{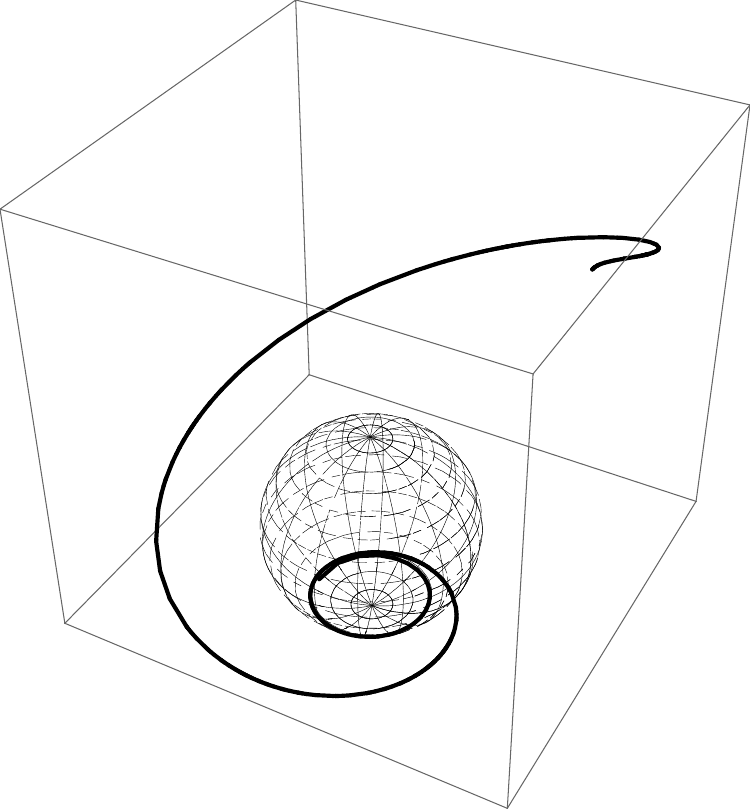}
\includegraphics[width=60mm]{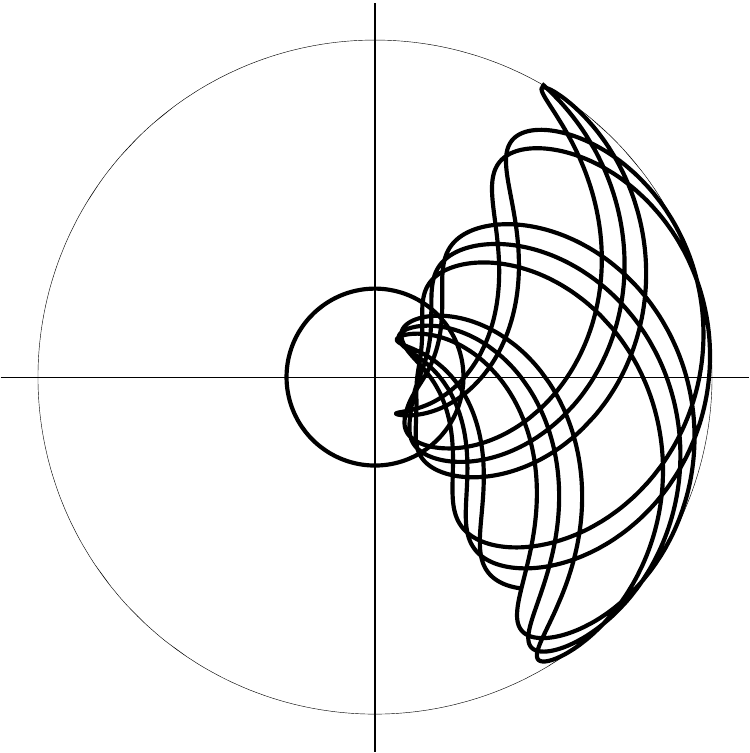}
\end{center}
\caption{Trajectory for a particle that fall to the horizon. Left figure for three-dimensional motion, and right figure is the motion in the $r$-$\theta$ plane, for $a=1.9$, $M=2$, $r_0=1$, $r_1=10$, $k=4$ and $L=2$.}
\label{o1}
\end{figure}

\begin{figure}[!h]
\begin{center}
\includegraphics[width=60mm]{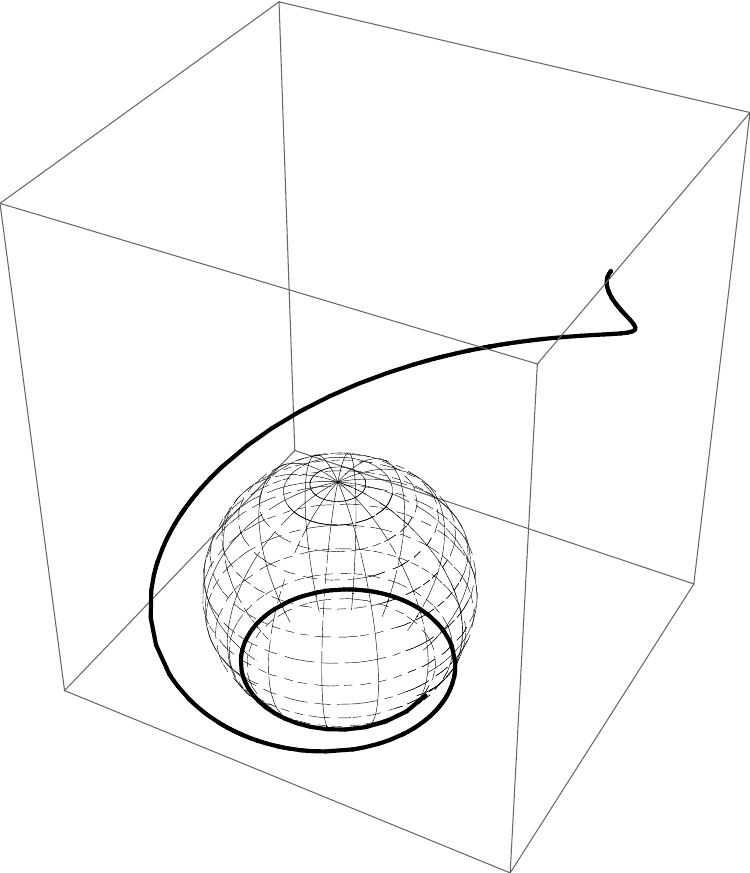}
\includegraphics[width=40mm]{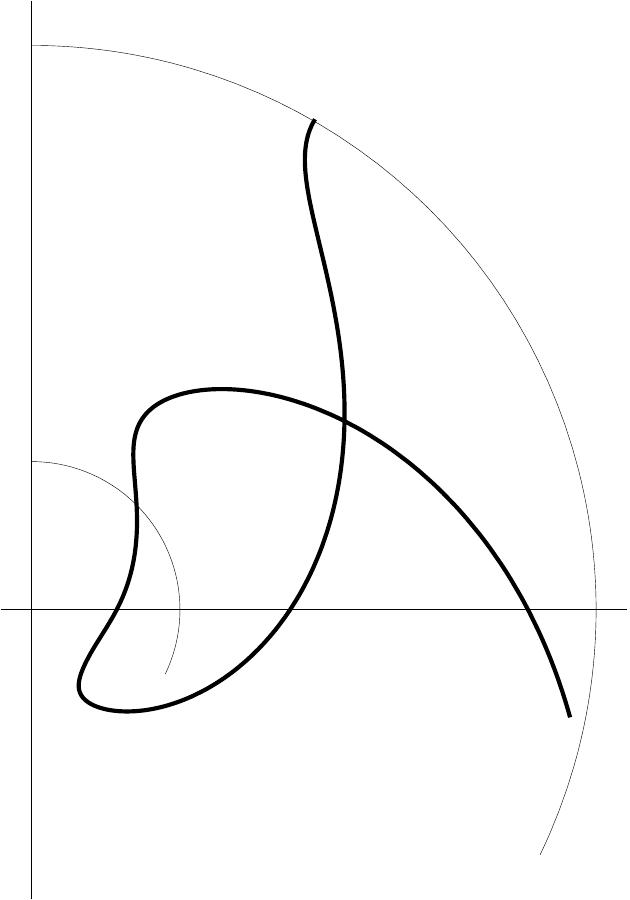}
\end{center}
\caption{Trajectory for a photon that fall to the horizon. Left figure for three-dimensional motion, and right figure is the projection of the orbit onto the x-z plane, for $a=1.9$, $M=2$, $r_0=1$, $\rho_1=10$, $k=4$, and $L=2$}
\label{o3}
\end{figure}

\begin{figure}[!h]
\begin{center}
\includegraphics[width=60mm]{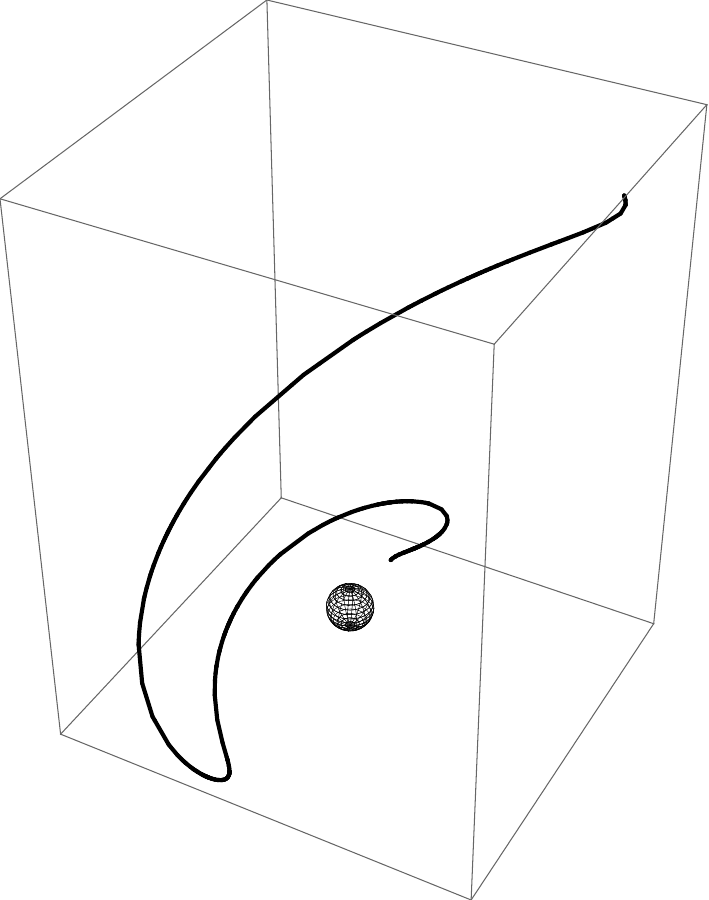}
\includegraphics[width=40mm]{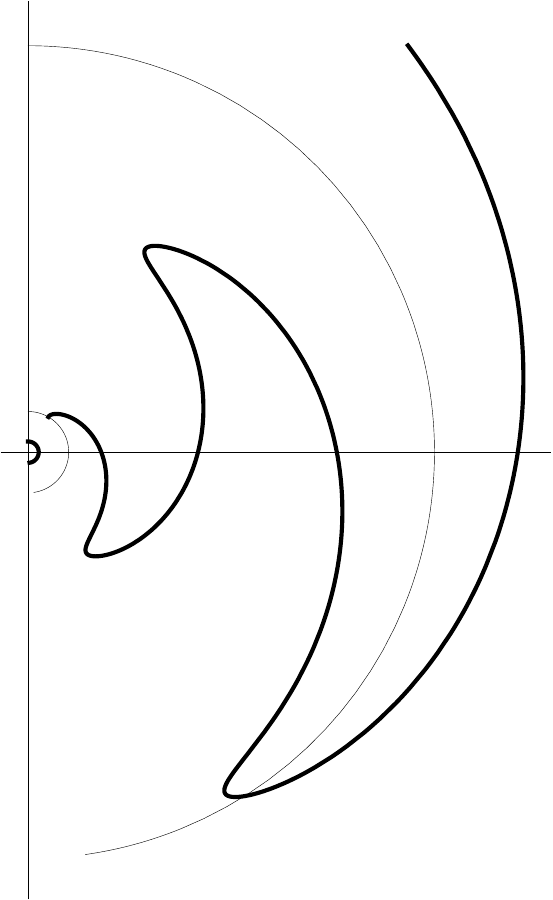}
\end{center}
\caption{Trajectory for a photon that escape to infinity. Left figure for three-dimensional motion, and right figure is the projection of the orbit onto the x-z plane, for $a=1.9$, $M=2$, $k=4$, $L=10$, and $\rho_1=10$.}
\label{o4}
\end{figure}

Now, in order to show the trajectory for particles and photons, we plot in Fig. \ref{o1} the trajectory of a particle that fall to the black hole (left figure) and the motion in the $r$-$\theta$ plane (right figure). Then, in Fig. \ref{o3} we plot the trajectory of a photon that fall to the black hole (left figure) and the motion in the $r$-$\theta$ plane (right figure). Finally, in Fig. \ref{o4} we plot the trajectory of a photon that escape to the infinity (left figure) and the motion in the $r$-$\theta$ plane (right figure).
\section{Summary}\label{conclusion}

In this paper we studied the motion of particles in a linear dilaton black hole background. We analyzed the motion of particles in the equatorial plane as well as in four dimensions analytically. Mainly, a qualitative analysis of the effective potential for null geodesics, shows that photons plunge into the horizon or escape to infinity,  they are not deflected. The study for massive particles shows that there are not confined orbits of first kind, like planetary or circular orbit. Therefore, the $r$-motion is not periodic, and it is not possible to define the orbital frequencies $\Omega_r$,  $\Omega_{\theta}$ and $\Omega_{\phi}$,
that determine
the perihelion shift and the Lense-Thirring effects,  which are defined as differences between these
orbital frequencies. Now, considering the $\theta$-motion, we observe that the polar coordinate is confined to oscillate between $\theta_1$ and  $\theta_2$.  The $\theta$-motion is periodic  with a
period
\begin{equation}
\omega_{\theta}=2\gamma(\theta_2)={2\pi\over k}~,
\end{equation}
and the corresponding frequency is $2\pi/\omega_{\theta}$. The secular accumulation rates of the angle $\phi$  is  given by:
\begin{equation}
Y_{\phi}^{(m)}={2\over\omega_{\theta}}\phi(\theta_2)= {2\over\omega_{\theta}}\left[ \varphi^{m}(\theta_2)+\varphi(\theta_2)\right] =
 {k\over\pi}\left[ \varphi^{m}(\theta_2)+\pi\right] ~,
\end{equation}
where $\varphi^{m}(\theta_2)= \varphi^{m}[r^m({\pi\over k})]$,
and $\gamma(\theta_2)={\pi\over k}$ .  The secular accumulation rates of the angle $\phi$  is   different for  particles and photons. It is worth to mention that another remarkable feature of the motion of photons is the existence of circular orbits of any radius. This occurs in the extremal case $(M=a)$  and for $L=Er_0$.

Also, we analyzed the collision of particles near the horizon and we studied the possibility that the dilaton black hole acts as a particle accelerator. We found that when one of the particles with critical angular momentum, which exists inside the outer horizon, collides with another  particle, it can  produce an arbitrarily high CM energy, in the non-extremal case. For the extremal case, we have found that the particles with critical angular momentum can only exist on the degenerate horizon. The above arguments show that the motion and collision of particles in the linear dilaton black hole has a similar behavior to the observed for the motion and collision of particles in the BTZ black hole \cite{Cruz:1994ir, Lake:2010bq, Yang:2012we, Tsukamoto:2017rrl}. Also, we showed that in the equatorial plane the string coupling depends on $r_0$ and it does not depend on $a$. Then we  observed that when the coupling $r_0$ is increasing, the  string coupling becomes stronger and this leads to a higher value of the critical angular momentum. Thus, a strong string coupling  requires the particle to have a high angular momentum in the collision process in order  the rotating linear dilaton black hole to act as a particle  accelerator.


\begin{acknowledgments}
This work was partially funded by the Comisi\'{o}n
Nacional de Ciencias y Tecnolog\'{i}a through FONDECYT Grant 11140674 (PAG) and by the Direcci\'{o}n de Investigaci\'{o}n y Desarrollo de la Universidad de La Serena (Y.V.). P. A. G. acknowledges the hospitality of the Universidad de La Serena and National Technical University of Athens and E. P.
and Y. V. acknowledge the hospitality of the Universidad Diego Portales where part of this work was carried out.
\end{acknowledgments}

\newpage

\end{document}